%Paper: hep-th/9205072
%From: Bert Schellekens <SCHELLEK%CERNVM.BITNET@pucc.princeton.edu>
%Date: Wed, 20 May 92 13:04:30 SET

\input phyzzx
% In addition to phyzzx, the `tables' macro is needed. all other
% macros are in this file.
%
\catcode`@=11
\def\space@ver#1{\let\@sf=\empty \ifmmode #1\else \ifhmode
   \edef\@sf{\spacefactor=\the\spacefactor}\unskip${}#1$\relax\fi\fi}
\def\attach#1{\space@ver{\strut^{\mkern 2mu #1} }\@sf\ }
\newtoks\foottokens
\newbox\leftpage \newdimen\fullhsize \newdimen\hstitle \newdimen\hsbody
\newif\ifreduce  \reducefalse
\def\almostshipout#1{\if L\lr \count2=1
      \global\setbox\leftpage=#1 \global\let\lr=R
  \else \count2=2
    \shipout\vbox{\special{dvitops: landscape}
      \hbox to\fullhsize{\box\leftpage\hfil#1}} \global\let\lr=L\fi}
\def\smallsize{\relax
\font\eightrm=cmr8 \font\eightbf=cmbx8 \font\eighti=cmmi8
\font\eightsy=cmsy8 \font\eightsl=cmsl8 \font\eightit=cmti8
\font\eightt=cmtt8
\def\eightpoint{\relax
\textfont0=\eightrm  \scriptfont0=\sixrm
\scriptscriptfont0=\sixrm
\def\rm{\fam0 \eightrm \f@ntkey=0}\relax
\textfont1=\eighti  \scriptfont1=\sixi
\scriptscriptfont1=\sixi
\def\oldstyle{\fam1 \eighti \f@ntkey=1}\relax
\textfont2=\eightsy  \scriptfont2=\sixsy
\scriptscriptfont2=\sixsy
\textfont3=\tenex  \scriptfont3=\tenex
\scriptscriptfont3=\tenex
\def\it{\fam\itfam \eightit \f@ntkey=4 }\textfont\itfam=\eightit
\def\sl{\fam\slfam \eightsl \f@ntkey=5 }\textfont\slfam=\eightsl
\def\bf{\fam\bffam \eightbf \f@ntkey=6 }\textfont\bffam=\eightbf
\scriptfont\bffam=\sixbf   \scriptscriptfont\bffam=\sixbf
\def\tt{\fam\ttfam \eightt \f@ntkey=7 }
\def\caps{\fam\cpfam \tencp \f@ntkey=8 }\textfont\cpfam=\tencp
\setbox\strutbox=\hbox{\vrule height 7.35pt depth 3.02pt width\z@}
\samef@nt}
\def\Eightpoint{\eightpoint \relax
  \ifsingl@\subspaces@t2:5;\else\subspaces@t3:5;\fi
  \ifdoubl@ \multiply\baselineskip by 5
            \divide\baselineskip by 4\fi }
\parindent=16.67pt
\itemsize=25pt
\thinmuskip=2.5mu
\medmuskip=3.33mu plus 1.67mu minus 3.33mu
\thickmuskip=4.17mu plus 4.17mu
\def\thinspace{\kern .13889em }
\def\negthinspace{\kern-.13889em }
\def\enspace{\kern.416667em }
\def\enskip{\hskip.416667em\relax}
\def\quad{\hskip.83333em\relax}
\def\qquad{\hskip1.66667em\relax}
\def\crr{\cropen{8.3333pt}}
\foottokens={\Eightpoint\singlespace}
\def\papersize{\SIZE\OFFSET\skip\footins=\bigskipamount}
\def\SIZE{\hsize=11.8truecm\vsize=17.5truecm}
\def\OFFSET{\voffset=-1.3truecm\hoffset=  .14truecm}
\def\attach##1{\space@ver{\strut^{\mkern 1.6667mu ##1} }\@sf\ }
\def\PH@SR@V{\doubl@true\baselineskip=20.08pt plus .1667pt minus .0833pt
             \parskip = 2.5pt plus 1.6667pt minus .8333pt }
\def\author##1{\vskip\frontpageskip\titlestyle{\tencp ##1}\nobreak}
\def\address##1{\par\kern 4.16667pt\titlestyle{\tenpoint\it ##1}}
\def\andaddress{\par\kern 4.16667pt \centerline{\sl and} \address}
\def\abstract{\vskip\frontpageskip\centerline{\twelverm ABSTRACT}
              \vskip\headskip }
\def\cases##1{\left\{\,\vcenter{\Tenpoint\m@th
    \ialign{$####\hfil$&\quad####\hfil\crcr##1\crcr}}\right.}
\def\matrix##1{\,\vcenter{\Tenpoint\m@th
    \ialign{\hfil$####$\hfil&&\quad\hfil$####$\hfil\crcr
      \mathstrut\crcr\noalign{\kern-\baselineskip}
     ##1\crcr\mathstrut\crcr\noalign{\kern-\baselineskip}}}\,}
\Tenpoint
}
\def\Smallsize{\smallsize\reducetrue
\let\lr=L
\hstitle=8truein\hsbody=4.75truein\fullhsize=24.6truecm\hsize=\hsbody
\output={
  \almostshipout{\leftline{\vbox{\makeheadline
  \pagebody\makefootline}}}\advancepageno
     }
\special{dvitops: landscape}
\def\makeheadline{
\iffrontpage\line{\the\headline}
             \else\vskip .0truecm\line{\the\headline}\vskip .5truecm \fi}
\def\makefootline{\iffrontpage\vskip  0.truecm\line{\the\footline}
               \vskip -.15truecm\line{\the\date\hfil}
              \else\line{\the\footline}\fi}
\paperheadline={
\iffrontpage\hfil
               \else
               \tenrm\hss $-$\ \folio\ $-$\hss\fi    }
\paperstyle}
%
%%%%%%%%%%%%%%%%%%%%%%%%%%%%%%%%%%%%%%%%%%%%%%%%%%%%%%%%%%%%%%%%%%%%%%%%
%
% Macros for Automatic Reference Numbering
%
%  All references are in a reference file in the following form:
%
%   \def\AAA{\rrr\AAA{This is reference AAA}}
%   \def\BBB{\rrr\BBB{This is reference BBB}}  ...etc
%
%  Then a reference is called as \BBB, and one may call it
%  as often as desired. Before the first reference occurs
%  one should read the reference file by adding \input (fn); the
%  filetype is assumed to be TEX. At the end of the paper an
%  ordered referencelist is produced by \refout.
%
%  Multiple references are called as \multref\AAA{\BBB\CCC\DDD...\ZZZ},
%  which yields e.g. [1-7].
%  Two references may also be called by \doubref\AAA\BBB, which
%  yields e.g. [5,6].
%
%  All PHYZZX routines remain intact.
%
%%%%%%%%%%%%%%%%%%%%%%%%%%%%%%%%%%%%%%%%%%%%%%%%%%%%%%%%%%%%%%%%%%%%
%
\newcount\referencecount     \referencecount=0
\newif\ifreferenceopen       \newwrite\referencewrite
\newtoks\rw@toks
\def\NPrefmark#1{\attach{\scriptscriptstyle [ #1 ] }}
\let\PRrefmark=\attach
\def\refmark#1{\relax\ifPhysRev\PRrefmark{#1}\else\NPrefmark{#1}\fi}
\def\refend{\refmark{\number\referencecount}}
\newcount\lastrefsbegincount \lastrefsbegincount=0
\def\refsend{\refmark{\count255=\referencecount
   \advance\count255 by-\lastrefsbegincount
   \ifcase\count255 \number\referencecount
   \or \number\lastrefsbegincount,\number\referencecount
   \else \number\lastrefsbegincount-\number\referencecount \fi}}
\def\refch@ck{\chardef\rw@write=\referencewrite
   \ifreferenceopen \else \referenceopentrue
   \immediate\openout\referencewrite=referenc.texauxil \fi}
%
% In \obeyendofline, we say `\let^^M=\relax
{\catcode`\^^M=\active % these lines must end with %
  \gdef\obeyendofline{\catcode`\^^M\active \let^^M\ }}%
%
% In \ignoreendofline, we say `\let^^M=\relax
{\catcode`\^^M=\active % these lines must end with %
  \gdef\ignoreendofline{\catcode`\^^M=5}}
{\obeyendofline\gdef\rw@start#1{\def\t@st{#1} \ifx\t@st\blankend%
\endgroup \@sf \relax \else \ifx\t@st\bl@nkend \endgroup \@sf \relax%
\else \rw@begin#1
\backtotext
\fi \fi } }
{\obeyendofline\gdef\rw@begin#1
{\def\n@xt{#1}\rw@toks={#1}\relax%
\rw@next}}
\def\blankend{}
{\obeylines\gdef\bl@nkend{
}}
\newif\iffirstrefline  \firstreflinetrue
\def\rwr@teswitch{\ifx\n@xt\blankend \let\n@xt=\rw@begin %
 \else\iffirstrefline \global\firstreflinefalse%
\immediate\write\rw@write{\noexpand\obeyendofline \the\rw@toks}%
\let\n@xt=\rw@begin%
      \else\ifx\n@xt\rw@@d \def\n@xt{\immediate\write\rw@write{%
        \noexpand\ignoreendofline}\endgroup \@sf}%
             \else \immediate\write\rw@write{\the\rw@toks}%
             \let\n@xt=\rw@begin\fi\fi \fi}
\def\rw@next{\rwr@teswitch\n@xt}
\def\rw@@d{\backtotext} \let\rw@end=\relax
\let\backtotext=\relax

\newdimen\refindent     \refindent=30pt
\def\refitem#1{\par \hangafter=0 \hangindent=\refindent \Textindent{#1}}
\def\REFNUM#1{\space@ver{}\refch@ck \firstreflinetrue%
 \global\advance\referencecount by 1 \xdef#1{\the\referencecount}}
\def\refnum#1{\space@ver{}\refch@ck \firstreflinetrue%
 \global\advance\referencecount by 1 \xdef#1{\the\referencecount}\refend}

\def\REF#1{\REFNUM#1%
 \immediate\write\referencewrite{%
 \noexpand\refitem{#1.}}%
\begingroup\obeyendofline\rw@start}
\def\ref{\refnum\?%
 \immediate\write\referencewrite{\noexpand\refitem{\?.}}%
\begingroup\obeyendofline\rw@start}
\def\Ref#1{\refnum#1%
 \immediate\write\referencewrite{\noexpand\refitem{#1.}}%
\begingroup\obeyendofline\rw@start}
\def\REFS#1{\REFNUM#1\global\lastrefsbegincount=\referencecount
\immediate\write\referencewrite{\noexpand\refitem{#1.}}%
\begingroup\obeyendofline\rw@start}
\newif\ifmref  %check multi ref
\newif\iffref  %check first ref
\def\xrefsend{\xrefmark{\count255=\referencecount
\advance\count255 by-\lastrefsbegincount
\ifcase\count255 \number\referencecount
\or \number\lastrefsbegincount,\number\referencecount
\else \number\lastrefsbegincount-\number\referencecount \fi}}
\def\xrefsdub{\xrefmark{\count255=\referencecount
\advance\count255 by-\lastrefsbegincount
\ifcase\count255 \number\referencecount
\or \number\lastrefsbegincount,\number\referencecount
\else \number\lastrefsbegincount,\number\referencecount \fi}}
\def\xREFNUM#1{\space@ver{}\refch@ck\firstreflinetrue%
\global\advance\referencecount by 1
\xdef#1{\xrefend}}
\def\xrefend{\xrefmark{\number\referencecount}}
\def\xrefmark#1{[{#1}]}
\def\xRef#1{\xREFNUM#1\immediate\write\referencewrite%
{\noexpand\refitem{#1 }}\begingroup\obeyendofline\rw@start}%
\def\xREFS#1{\xREFNUM#1\global\lastrefsbegincount=\referencecount%
\immediate\write\referencewrite{\noexpand\refitem{#1 }}%
\begingroup\obeyendofline\rw@start}
\def\rrr#1#2{\relax\ifmref{\iffref\xREFS#1{#2}%
\else\xRef#1{#2}\fi}\else\xRef#1{#2}\xrefend\fi}
\def\multref#1#2{\mreftrue\freftrue{#1}%
\freffalse{#2}\mreffalse\xrefsend}
\def\doubref#1#2{\mreftrue\freftrue{#1}%
\freffalse{#2}\mreffalse\xrefsdub}
\referencecount=0
\def\refbreak{\hfil\penalty200\hfilneg}
\def\paperstyle{\papers}
\paperstyle   %  This is the default
%
%%%%%%%%%%%%%%%%%%%%%%%%%%%%%%%%%%%%%%%%%%%%%%%%%%%%%%%%%%%%%%%%%%%%
%
%  Local, ie. site-dependent macros.
%
\def\slacpub{\afterassignment\slacp@b\toks@}
\def\slacp@b{\edef\n@xt{\Pubnum={CERN--TH.\the\toks@}}\n@xt}
\let\pubnum=\slacpub
\expandafter\ifx\csname eightrm\endcsname\relax
    \let\eightrm=\ninerm \let\eightbf=\ninebf \fi

\font\seventeencp=cmcsc10 scaled\magstep3

\newif\ifCONF \CONFfalse
\newif\ifBREAK \BREAKfalse
\newif\ifsectionskip \sectionskiptrue

%
%%%%%%%%%%%%%%%%%%%%%%%%%%%%%%%%%%%%%%%%%%%%%%%%%%%%%%%%%%%%%%%%%%%%%%%%
%           FORMATS
%%%%%%%%%%%%%%%%%%%%%%%%%%%%%%%%%%%%%%%%%%%%%%%%%%%%%%%%%%%%%%%%%%%%%%%%
%
% Nuclear Physics Proceedings Format
%
%
\def\NuclPhysProc{
\let\lr=L
\hstitle=8truein\hsbody=4.75truein\fullhsize=21.5truecm\hsize=\hsbody
\hstitle=8truein\hsbody=4.75truein\fullhsize=20.7truecm\hsize=\hsbody
\output={
  \almostshipout{\leftline{\vbox{\makeheadline
  \pagebody\makefootline}}}\advancepageno
     }
\def\papersize{\SIZE\OFFSET\skip\footins=\bigskipamount}
\def\SIZE{\hsize=10.0truecm\vsize=27.0truecm}
\def\OFFSET{\voffset=-1.4truecm\hoffset=-2.40truecm}
\def\makeheadline{
\iffrontpage\line{\the\headline}
             \else\vskip .0truecm\line{\the\headline}\vskip .5truecm \fi}
\def\makefootline{\iffrontpage\vskip  0.truecm\line{\the\footline}
               \vskip -.15truecm\line{\the\date\hfil}
              \else\line{\the\footline}\fi}
\paperheadline={\hfil}
\paperstyle}
%
%
% World Scientific Format
%
%

\paperstyle
%
%
% Reprint Volume Format
%
%
\def\ReprintVolume{\smallsize
\def\papersize{\hsize=18.0truecm\vsize=25.1truecm\voffset -.73truecm
    \hoffset -.65truecm\skip\footins=\bigskipamount
    \normaldisplayskip= 20pt plus 5pt minus 10pt}
\paperstyle\baselineskip=.425truecm\parskip=0truecm
\def\makeheadline{
\iffrontpage\line{\the\headline}
             \else\vskip .0truecm\line{\the\headline}\vskip .5truecm \fi}
\def\makefootline{\iffrontpage\vskip  0.truecm\line{\the\footline}
               \vskip -.15truecm\line{\the\date\hfil}
              \else\line{\the\footline}\fi}
\paperheadline={
\iffrontpage\hfil
               \else
               \tenrm\hss $-$\ \folio\ $-$\hss\fi    }
\def\sectionfont{\bf}    }
%
%
% CERN-preprint format (default)
%
%
\def\SIZE{\hsize=15.73truecm\vsize=23.11truecm}
\def\OFFSET{\voffset=0.4truecm\hoffset=-0.88truecm}
\def\papersize{\SIZE\OFFSET\skip\footins=\bigskipamount
\normaldisplayskip= 35pt plus 3pt minus 7pt}
\def\CERN{\address{{\sl CERN, 1211 Geneva 23, Switzerland\
\phantom{XX}\ }}}
\Pubnum={\rm CERN$-$TH.\the\pubnum }
\def\title#1{\vskip\frontpageskip\vskip .50truein
     \titlestyle{\seventeencp #1} \vskip\headskip\vskip\frontpageskip
     \vskip .2truein}
\def\author#1{\vskip .27truein\titlestyle{#1}\nobreak}

\def\p@bblock{\begingroup \tabskip=\hsize minus \hsize
   \baselineskip=1.5\ht\strutbox \topspace-2\baselineskip
   \halign to\hsize{\strut ##\hfil\tabskip=0pt\crcr
   \the \Pubnum\cr}\endgroup}
\def\makefootline{\iffrontpage\vskip .27truein\line{\the\footline}
                 \vskip -.1truein\line{\the\date\hfil}
              \else\line{\the\footline}\fi}
\paperfootline={\iffrontpage
 \the\Pubnum\hfil\else\hfil\fi}
\paperheadline={
\iffrontpage\hfil
               \else
               \twelverm\hss $-$\ \folio\ $-$\hss\fi}
%
% Journal abbreviations
%
\def\nup#1({\refbreak\ Nucl.\ Phys.\ $\underline {B#1}$\ (}
\def\plt#1({\refbreak\ Phys.\ Lett.\ $\underline  {#1}$\ (}
\def\cmp#1({\refbreak\ Commun.\ Math.\ Phys.\ $\underline  {#1}$\ (}
\def\prp#1({\refbreak\ Physics\ Reports\ $\underline  {#1}$\ (}
\def\prl#1({\refbreak\ Phys.\ Rev.\ Lett.\ $\underline  {#1}$\ (}
\def\prv#1({\refbreak\ Phys.\ Rev. $\underline  {D#1}$\ (}
\def\und#1({            $\underline  {#1}$\ (}
%
% Line breaks
%

\def\rB{\hfil\penalty1000\hfilneg}
%
% Hyphenations
%
\hyphenation{sym-met-ric anti-sym-me-tric re-pa-ra-me-tri-za-tion
Lo-rentz-ian a-no-ma-ly di-men-sio-nal two-di-men-sio-nal}
%
%
% Miscellaneous macros
%
%

%

%%

%

%
\def\boxit#1{\vbox{\hrule\hbox{\vrule\kern3pt
\vbox{\kern3pt#1\kern3pt}\kern3pt\vrule}\hrule}}
\message{ by V.K, W.L and A.S}
\catcode`@=12
\paperstyle
%%% REFERENCES %%%
\def\FHPV{\rrr\FHPV{
P.~Forgacs, L.~Palla,  Z.~Horvath and P.~Vecserny\'es,
\nup308 (1988) 477.}}
\def\DrFu{\rrr\DrFu{J.~Fuchs and P. van Driel, Lett. Math. Phys. 23
 (1991) 11.}}
\def\Lee {\rrr\Lee {J.~Leech, Canadian Journal of Mathematics
\und 19 (1967) 251.}}
\def\Sie {\rrr\Sie {C.~Siegel, Ann.~Math.~\und  36 (1935) 527.}}
\def\CoS{\rrr\CoS{
J.~Conway and M.~Sloane, Journal of Number Theory \und 15 (1982) 83; \rB
Europ.~J.~Combinatorics 3 (1982) 219.}}
\def\FQS{\rrr\FQS{D.~Friedan, Z.~Qiu and S.~Shenker,
\prl52  (1984) 1575.}}
\def\ScW {\rrr\ScW {A.N.~Schellekens
and N.P.~Warner, \plt B177 (1986) 317;
\plt B181 (1986) 339;     \nup287 (1987) 317.}}
\def\FGK {\rrr\FGK {G.~Felder, K.~Gawedzki and A.~Kupainen,
\cmp 117 (1988) 127.}}
\def\DiH{\rrr\DiH{
L.~Dixon and J.~Harvey, \nup274 (1986) 93.}}
\def\LLSA{\rrr\LLSA{
W.~Lerche, D.~L\"ust and A.N.~Schellekens, \plt B181 (1986) 71;
Erratum, \plt B184 (1987) 419.}}
\def\SchH{\rrr\SchH{
A.N.~Schellekens,
\plt B277 (1992) 277.}}
\def\KaPe{\rrr\KaPe{
V.G.~Kac and D.H.~Peterson, Adv.~in Math.~53 (1984) 125.}}
\def\Nie{\rrr\Nie{
H.~Niemeier, Journal of Number Theory \und 5 (1973) 142.}}
\def\GKO{\rrr\GKO{
P.~Goddard, A.~Kent and D.~Olive,
\plt152B (1985) 88;\rB \cmp103 (1986) 105.}}
\def\CoSu{\rrr\CoSu{F.~Bais and
P.~Bouwknegt,
\nup279 (1987) 561; \rB A.N.~Schellekens and N.P.~Warner,
\prv34 (1986) 3092.}}
\def\BPZ {\rrr\BPZ {A.~Belavin, A.~Polyakov and A.~Zamolodchikov,
 \nup241 (1984) 333.}}
\def\AGMV{\rrr\AGMV{L.~Alvarez-Gaum\'e, P.~Ginsparg, G.~Moore and
                    C.~Vafa, \plt B171 (1986) 155.}}
\def\KLTE{\rrr\KLTE{H.~Kawai, D.~Lewellen and S.~Tye,
\prv34 (1986) 3794.}}
\def\ScYc{\rrr\ScYc{A.N.~Schellekens and S.~Yankielowicz,
%{\it Curiosities at $c = 24$},
\plt B226 (1989) 285.}}
\def\BB  {\rrr\BB  {L.~Dixon, P.~Ginsparg and J.~Harvey,
%beauty and the beast
\cmp119 (1988) 221.}}
\def\God {\rrr\God {P.~Goddard, {\it Meromorphic Conformal Field Theory},
preprint DAMTP-89-01.}}
\def\EVeA{\rrr\EVeA{E.~Verlinde,
%{\it Fusion Rules and Modular
%transformations in 2D Conformal Field Theory}
\nup300 (1988) 360.}}
\def\ZamA{\rrr\ZamA{A.~Zamolodchikov, Theor.\ Math.\ Phys.\
\und 65 (1986) 1205.}}
\def\LSW {\rrr\LSW {W.~Lerche, A.~Schellekens and N.~Warner,
%{\it Ghost Triality and Superstring Partition Functions},
\plt B214 (1988) 41.}}
\def\AhWa{\rrr\AhWa{C.~Ahn and M.~Walton,\plt B223 (1989) 343.}}
\def\Gin {\rrr\Gin {P.~Ginsparg, {\it Informal String Lectures},
 in the Proceedings of the U.K.~Institute for Theoretical
 High Energy Physics, Cambridge, 16 Aug.~- 5 Sept.~1987, Harvard preprint
 HUTP-87/A077 (1987).}}
\def\BoNa{\rrr\Bona{P.~Bouwknegt and W.~Nahm,
\plt B184 (1987) 359.}}
\def\DVVV{\rrr\DVVV{R.~Dijkgraaf, C.~Vafa, E.~Verlinde and
H.~Verlinde,
%{\it The Operator Algebra of Orbifold Models},
\cmp 123 (1989) 16.}}
\def\Kir {\rrr\Kir {E.~Kiritsis, \nup324 (1989) 475.}}
%(Fuchsian differential eqns)
%
\def\ScYb{\rrr\ScYb{
A.N.~Schellekens and S.~Yankielowicz,
%{\it Modular Invariants from simple currents: An explicit proof},
\plt B227 (1989) 387.}}
\def\ScYA{\rrr\ScYA{
A.N.~Schellekens and S.~Yankielowicz,
\nup 327 (1989) 673; \rB
\plt B227 (1989) 387.}}
\def\CIZ {\rrr\CIZ {
A.~Cappelli, C.~Itzykson     and J.-B.~Zuber,
\nup280 (1987)  445;\rB \cmp113 (1987) 1.}}
\def\Ber {\rrr\Ber {
D.~Bernard, \nup288 (1987) 628.}}
\def\ALZ {\rrr\ALZ {D.~Altschuler, J.~Lacki and P.~Zaugg, \plt B205
(1988) 281.}}
\def\Alig{\rrr\Alig{K.~Intriligator,
%{\it Bonus Symmetry in Conformal Field Theory}, \rB
\nup 332 (1990) 541.}}
\def\Fuch{\rrr\Fuch{J.~Fuchs,
\cmp136 (1991) 345.}}
\def\BeBT{\rrr\BeBT {B.~Gato-Rivera and A.N.~Schellekens,
%{\it Complete Classification of Simple Current Modular Invariants
%for $({\bf Z}_p)^k$}
\cmp 145 (1992)  85.}}
\def\DGM {\rrr\DGM  {L.~Dolan, P.~Goddard and P.~Montague,
\plt B236 (1990) 165.}}
\def\Ver {\rrr\Ver  {D.~Verstegen, \nup 346 (1990) 349;
\cmp 137 (1991) 567.}}
\def\LFM {\rrr\LFM  {I.~Frenkel, J.~Lepowsky and J.~Meurman,
Proc. Natl. Acad. Sci. U.S.A. 81, 32566 (1984).}}
\def\KaWa{\rrr\KaWa{V.Kac and M. Wakimoto, Adv. Math. 70 (1988) 156.}}
\def\Walt{\rrr\Walt{M. Walton, \nup322 (1989) 775.}}
\def\ABI {\rrr\ABI {D. Altschuler, M. Bauer and C. Itzykson,
\cmp132 (1990) 349.}}
\def\Pat {\rrr\Pat {J. Math. Phys. 17 (1976) 1972;
S. Okubo and J. Patera, J. Math. Phys. 24 (1983)
2722; 25 (1984) 219. }}
\hfuzz= 6pt
\def\Zbf{{\bf Z}}
\def\mod{{\rm ~mod~}}

\def\N{{\cal N}}
\def\X{{\cal X}}
\def\ie{i.e.}
\def\eg{e.g.}

\pubnum={6478/92}
\date{April 1992 --- hep-th/9205072  }
\pubtype{CRAP}
\titlepage
\title{Meromorphic c=24 Conformal Field Theories}
\author{A. N. Schellekens}
\vskip 0.3truein
\CERN
\vskip 0.6truein
\abstract
Modular invariant conformal field theories with just one primary
field and central charge $c=24$ are considered. It has been shown
previously that if the chiral algebra
of such a theory contains spin-1 currents, it is either the Leech
lattice CFT, or it contains a Kac-Moody sub-algebra with total
central charge 24. In this paper all meromorphic
modular invariant combinations
of the allowed Kac-Moody combinations are obtained. The result
suggests the existence of 71 meromorphic $c=24$ theories, including
the 41 that were already known.
\endpage
%\paperheadline={\hfil}  \paperstyle
%~~~~~~~~~~~~~~~~~~~~~~~~~~~~~~~~~
%\endpage
%\paperheadline={
%\iffrontpage\hfil
%               \else
%         \twelverm\hss $-$\ \folio\ $-$\hss\fi    } \paperstyle
\chapternumber=0
\pagenumber=1

\chapter{Introduction}

A conformal field theory is characterized by two algebraic structures:
the chiral algebra and the fusion algebra. The chiral algebra consists
at least of the Virasoro algebra, which in general is extended by
other operators of integer conformal weight. The representations or
primary fields of the chiral algebra obey a set of fusion rules,
determining which primary fields can appear in the operator product of
two such fields. In general, both the chiral algebra and the
set of fusion rules are non-trivial.

If one is interested in classifying conformal field theories, it seems
natural to start with the simplest ones. For example, one might consider
theories in which either the chiral algebra or the fusion algebra is as
simple as possible.
Theories of the former kind form the ``minimal series'' \BPZ\
(whose chiral algebra consists
{\it a priori} only of the Virasoro algebra) and
have been classified
completely \multref\FQS{\GKO\CIZ}.
The theories with the simplest possible set of fusion rules are those
with only one primary field $\bf 1$, and a fusion rule ${\bf1}\times
{\bf 1}={ \bf 1}$. In such theories the entire non-trivial structure
resides
obviously in the chiral algebra.

Theories of this kind have extremely simple modular transformation
properties \EVeA. The identity is self-conjugate, and hence
the charge conjugation matrix $C$ must be equal to $1$.
Therefore
$S=\pm 1$ and the identity character $\X(\tau)$ satisfies
$\X(-{1\over\tau})=\pm \X(\tau)$. Choosing $\tau=i$, and noting that
the character is a polynomial with positive coefficients in
$q=e^{2\pi i \tau}$ so that $\X(i)\not=0$, we see that $S=1$.
Furthermore,
since $(ST)^3=C$, we find  $T=e^{2k\pi i\over 3}$, $k\in \Zbf$.
Since
$T=e^{2\pi i (h-{c\over 24})}$ and $h=0$ it follows that $c$ must be a
multiple of $8$.
The one-loop partition function of such a theory is simply $\X \X^*$,
where $\X$ is the character of the only representation of the theory.
If $c$ is a multiple of 24 the character itself is a modular invariant
partition function, and one can
consider a corresponding purely chiral conformal
field theory. In such a theory all correlation functions are
meromorphic, and hence these
theories have been called {\it meromorphic}
conformal field theories in
\God\rlap.\foot{This terminology is in fact
used in a
broader sense in \God. Throughout this paper we use the adjective
``meromorphic'' to indicate ``one-loop modular invariant meromorphic'',
or, in the terminology of \God,
``bosonic self-dual meromorphic".}

Examples of conformal field theories with just one character are
easy to construct. Consider
$N$ free chiral bosons whose momenta lie on an even self-dual
Euclidean lattice. The one-loop character $\X(\tau)$
is simply the lattice
partition function divided by $N$ $\eta$-functions.
Using Poisson resummation it is easy to show that this function
transforms into itself (up to phases) under both $S$ and $T$. Hence this
character transforms as a one-dimensional representation of the
modular group, and $c=N$ must be a multiple of 8, which indeed
is necessary in order to have an even self-dual Euclidean lattice.

Even self-dual Euclidean lattices have been classified
for dimensions 8 (the
root lattice of the Lie algebra $E_8$), 16
(the root lattice of $E_8 \times E_8$
and that of $D_{16}$ with the addition of a spinor weight), and 24
(the 24 Niemeier lattices \Nie). Any of these lattices defines a
distinct conformal field theory, and it is natural to ask whether
this exhausts the list of meromorphic conformal field theories in these
dimensions.

For $c=8$ and $c=16$ it is easy to see that this must be true. If there
were any meromorphic theories one could use them instead of $E_8$ or
$E_8\times E_8$ in the construction of the heterotic string. In
particular one could construct new modular invariant supersymmetric
heterotic strings, to which the relation
between modular invariance and cancellation of chiral anomalies
of \ScW\ (easily generalized to higher level,
see \Gin)
would apply.
Hence any such meromorphic theory would manifest
itself in the gravitational anomaly (and the gauge anomaly if there
are gauge fields) of the field theory.
Since only the two gauge groups $E_8\times E_8$
and $SO(32)$ were found to be allowed we know that no such theory
can exist.

This argument does not apply to the meromorphic $c=24$ theories.
Indeed, several theories are already known that cannot be described by
free bosons on Niemeier lattices. The first example is the
``monster module'' \LFM, a meromorphic $c=24$ theory without
any spin-1 operators, and which therefore lacks the
24 bosonic operators $\partial X$ that are present in any lattice
theory. This theory can be obtained by
applying
an orbifold $\Zbf_2$-twist $X \to -X$
to one of the 24 Niemeier lattice CFTs \doubref\BB\DGM.
Such a twist does not directly yield a meromorphic CFT, but a
theory with four primary fields
$E$, $O$, $\sigma_{\rm O}$ and $\sigma_{\rm E}$, with
spins 0, 1, ${3\over 2}$ and $2$ respectively.
The corresponding states
form respectively the odd and even states of the original Hilbert space,
and the odd and even states of the twisted Hilbert space.
The fusion rules
of these primary fields may be found in \DVVV.
In particular one has
$O^2=\sigma_{\rm O}^2=\sigma_{\rm E}^2=E$,
showing that all of them are simple currents.
Hence if they
have integral spin they can be put into the chiral algebra. Putting $O$
into the chiral algebra projects out the twist fields $\sigma$ and
$\tau$, and gives us back the original lattice theory, with
character $\X=\X_E+\X_O$.
The more
interesting possibility is to put
$\sigma_{\rm E}$ into the chiral algebra.
This projects out $J$ and $\sigma$, and gives a new, meromorphic
CFT whose character is $X_{E}+\X_{\sigma_{\rm E}}$.
The number of spin-1 operators in this new theory is easy to compute \BB,
and is given by
${\cal N}_{\rm twisted}={1\over2} {\cal N}_{\rm lattice} -12$,
where ${\cal N}$ denotes the number
of spin-1 operators in the chiral algebra.

This twisting procedure can be applied to any of the 24 Niemeier
lattices.
Although it always yields a different theory, that theory may itself be
another lattice theory. In \God\ it was shown in a very elegant way,
using
the theory of codes, that in 15 cases one obtains something new. One of
those 15 is the monster module, obtained from the Leech lattice CFT,
which
has ${\cal N}=24$. Thus we know now altogether
39 meromorphic $c=24$
theories.

The existence of one more theory can be inferred \ScYc\
from the existence
of a ten-dimensional (non-supersymmetric) heterotic string theory with
Kac-Moody algebra $E_{8,2}$ \KLTE\
(here and in the following $X_{m,n}$ denotes
an untwisted Kac-Moody algebra of type $X$, rank $m$ and level $n$).
In \LLSA\ a mapping from ten-dimensional heterotic strings
constructed
by means of the covariant lattice construction (or free complex fermions)
to a subset of the
Niemeier lattices was described.
This produced an easy classification of
all ten-dimensional heterotic strings with a rank-16 gauge group.
However,
the same map takes {\it any}
ten-dimensional heterotic string to some meromorphic $c=24$ theory.
The $E_{8,2}$ theory does not have a rank-16 gauge group and
cannot be described by a lattice, but it is still mapped to
some $c=24$ theory.
This theory
has 384 spin-1 currents forming a $B_{8,1}E_{8,2}$ Kac-Moody algebra,
which does not correspond to any twisted or untwisted
Niemeier theory. Hence it must be a new item on the growing
list of meromorphic $c=24$ theories.

The last example known to us before starting this work was found
more or less accidentally, as a result of a computer search for integer
spin extensions of Kac-Moody algebras \ScYA,\ScYc.
A modular invariant of $F_{4,6}$
emerged that was neither a simple current invariant nor a
conformal embedding. Although this new theory is
not a meromorphic theory, it turned out
that its six characters could be combined with the
six characters of $A_{2,2}$, so that a meromorphic theory was formed with
${\cal N}=60$. A twisted Niemeier theory exists with the same
number of spin-1 currents, but that theory has a Kac-Moody symmetry
$(C_{2,2})^6$, and is thus clearly different. This brings the total
so far to 41.

The goal of this paper is to complete the list of meromorphic
$c=24$ theories. This goal will indeed be achieved, but under three
additional assumptions. First of all
our methods require that ${\cal N}\not=0$.
It has been conjectured that there is just one theory with
${\cal N}=0$; for counting purposes this
will be assumed to be true in the following. Secondly, we will
assume that the chiral algebra is generated by a finite number of
currents. This is indeed true for all unitary rational conformal
field theories we know, and it might be possible to derive this
rather than assume it.
Finally, we will not really construct conformal field theories, but
modular invariant combinations of Kac-Moody characters. It will
be shown in the next section that if ${\cal N}\not=0$, then the chiral
algebra contains a spin-1 algebra with a total central charge 24. Hence
the partition function of any such theory must be a modular invariant
combination of Kac-Moody characters, and these combinations will be
enumerated completely. There are 69, not including the Leech lattice
and the monster module.
Barring the unlikely possibility of having
two or more distinct conformal field theories per combination (which
in any case must all have the same representation content in each
excitation level), this limits the set of possible distinct
$c=24$ theories to 71.
It remains to be proved that a conformal field theory
corresponding to all of these 71 partition functions actually exists.
In particular, one would like to write down the operator product
algebra of the set of the higher spin fields appearing in the
partition function. This is simply an example of
the familiar problem of writing down operator
products for non-diagonal theories. Methods to address this problem
exist and have been applied to various examples, but they are rather
laborious, and will not be pursued here.
The existence of non-diagonal modular invariant partition functions
requires a large number of conditions to be satisfied, and it is
difficult to believe that this would be a mere coincidence without
having the significance it strongly suggests.
For this reason we conjecture that a meromorphic
$c=24$ theory exists for any of the new partition functions.

Explicit constructions exist for the twisted and untwisted Niemeier
theories as well as the $B_{8,1}E_{8,2}$ theory (which can be
built out of real fermions), but for the second example
of \ScYc\ only the
modular invariant character combination is known.

There are several motivations for attempting to classify the
$c=24$ meromorphic conformal field theories. Originally our interest
in this problem was related to the aforementioned relation between
this classification and that of ten-dimensional
heterotic strings. Indeed, from a list of the $c=24$ theories one
can obtain a list of all $d=10$ heterotic strings by simply
looking for all possible embeddings of $D_{8,1}$.
Several $d=10$ heterotic strings have been constructed by various
methods \KLTE, \doubref\DiH\AGMV,
(in particular orbifolding, fermionic and lattice constructions),
but none of these methods has any claim to completeness.
In a recent paper we have proved completeness for $d=10$
heterotic strings
from
a partial classification of the meromorphic $c=24$ theories \SchH.
This work made it clear that with some more effort a complete
classification of the latter should be possible.
Although at present our main motivation for completing the
classification is
just curiosity, there are many
interesting facts related to $c=24$ that
suggest
these
theories might have their r\^ole to play (for example in connection
with the
bosonic string or the intriguing even
self-dual Lorentzian
lattice $\Gamma_{25,1}$,
or in connection with the Monster group).
In fact, we hoped
that the complete list might reveal an underlying structure that
was not apparent from the partial list, but this hope has not been
fulfilled so far.

Finally, the new solutions will provide us with interesting
information about a class of Kac-Moody modular invariants that is still
not understood at all. There are three known methods for
constructing in a systematic way extensions of the chiral algebra
of Kac-Moody algebras: simple currents, conformal embeddings and
rank-level duality. Simple currents \ScYA\
yield generalizations of the
$D$-invariants of $SU(2)$\rlap.\foot{In the special case of Kac-Moody
algebras, most of these invariants can be obtained by orbifolding
with respect to certain
extended Dynkin diagram automorphisms, which form
a group isomorphic to the center of the Lie algebra
\multref\Ber{\ALZ\FGK\AhWa}.
The only
exception \doubref\Fuch\FHPV\
is $E_{8,2}$, which has a simple current that can yield extensions
of the chiral algebra in certain tensor products.} Conformal embeddings
$H\subset G$ \CoSu\
imply extensions of the chiral algebra of $H$ to that
of $G$ \BoNa.
They can be recognized by the presence of spin-1 currents
in the extension.
Sometimes this method can be applied to embeddings $H_1 \times H_2
\subset G$ to infer the existence of invariants of $H_2$ from those
of $H_1$. Rank-level duality \multref\KaWa{\Walt\ABI\Ver}
can be viewed as a
special case
of this, and implies relations between the modular group
representations and invariants of the pairs  $SU(n)$ level $k$
$\leftrightarrow$ $SU(k)$ level $n$,
$C_{n,k}$ $\leftrightarrow$ $C_{k,n}$ and
$SO(n)$ level $k$ $\leftrightarrow$ $SO(k)$ level $n$.

Unfortunately this is not sufficient to obtain all extensions
of Kac-Moody algebras. Only one genuine
exception (which cannot be obtained by any combination of these
methods) was known so far, namely the $F_{4,6}$ invariant of \ScYc\
(note that we are not considering fusion rule automorphisms here).
The list of meromorphic $c=24$ theories yields several additional
examples.

The methods we used can be summarized as follows.
The starting point of the classification is the fact, mentioned
above, that the $c=8$ and $c=16$ theories can be classified using
anomaly cancellation in superstring theory. For $c=24$ we do not
use chiral anomalies of some string theory, but consider directly
the same trace-identities that imply anomaly cancellation in
superstring theory \ScW, and which contain in fact far more information.
This will be explained in the next section (some of the results have
already appeared in \SchH). The crucial observation is that the spin-1
currents of any $c=24$ theory must form a Kac-Moody algebra with
$c=24$, or a product of 24 $U(1)$'s.
(Note that this property does not hold for $c>24$; a trivial
counter-example is the monster module tensored with $E_{8,1}$. The
trouble is that there could be non-trivial counter-examples as well.)
Furthermore all Kac-Moody algebras appearing
in a given theory must have the same ratio $g/k$ (where $g$
is the dual Coxeter number and $k$ the level), and
$g/k$ can be computed from $\N$, the
total number of spin-1 currents. There are only 221 solutions to
these three conditions.

This is a very small number in comparison with the number of ways of
writing 24 as a sum of central charges of Kac-Moody algebras (not to
mention rational $U(1)$'s). Since this simple argument gets us so
close to the final answer, it is worthwhile to try to continue.
{}From here on, further reduction of the number of solutions is
considerably harder, though.

The next step is to use higher trace identities to rule out
accidental solutions. This is a fairly laborious task, but
one is finally left with 69 Kac-Moody combinations for which one
or more candidates for the
second excited level (with 196884 elements) exist that satisfy
all trace identities.
Now we consider directly the modular invariance conditions for the
remaining candidates. This looks hopeless at first, since the total
number of primary fields can be huge (for example $5^{12}$), and often
the number of integer spin fields is much too large as well. Here simple
currents come to the rescue. In many cases, we can conclude from the
already known results at the second level that certain simple currents
of spin 2 are present in the chiral algebra. This implies that some
primary fields are projected out, and the remaining ones are organized
into orbits. Each independent simple current of order $N$ reduces the
number of primary fields by a factor of $N^2$ (ignoring fixed points).
This makes it possible to find the solution. Indeed, for all of the
69 combinations for which a second-level solution exists we find
precisely one modular invariant partition function (up to
outer automorphisms).

\chapter{Trace identities}

In this section we construct the character-valued partition functions
for (unitary, modular invariant)
meromorphic conformal field theories with
central charge $8n$, $n\in \Zbf$,
and derive the resulting trace identities
for $c=24$.

Consider a unitary meromorphic theory with spin-1 operators,
\ie\ $\N\not=0$. Then these operators must have the following
operator product \ZamA\God
$$ J^a(z)J^b(w)={k\delta^{ab} \over (z-w)^2 } + {1\over z-w}
  i f^{abc}J^c(w) +  \hbox{ finite terms} \ . $$
If for some label $a$ all $f^{abc}$
vanish, then $J^a$ generates a $U(1)$ factor. Otherwise the
coefficients $f^{abc}$ must be structure constants of a semi-simple
Lie algebra, and the current algebra is then a Kac-Moody algebra.
The central charge
of the Sugawara energy momentum tensor of this algebra is smaller than
or equal to the total central charge. The states in the theory are
transforming in certain representations of the zero-mode algebra, and
this allows us to write down a ``character-valued'' partition function
containing this information:
$$ P(q,\vec F)=\Tr e^{\vec F\cdot\vec J_0} q^{L_0-c/24} \ ,\eqn\First$$
where $\vec F\cdot \vec J\equiv\sum_a F^a J^a      $,
and $F^a$ is a set of real coefficients. (In anomaly applications
$F^a$ would be a Yang-Mills two-form, but this will not be needed here.)

In general, the Kac-Moody algebra
generated by the spin-1 current consists
of several simple factors, and the partition function can be expressed in
terms of the characters $\X^{\ell}_{i_{\ell}}$ of the ${\ell}^{\rm th}$
factor and an unknown function without spin-1 contributions:
$$ P(q,\vec F_1,\ldots,\vec F_L)
=\sum_{i_1,\ldots,i_L}
\X^1_{i_1}(q,\vec F_1) \ldots \X^L_{i_L}(q,\vec F_L)
                 \X_{i_1,\ldots i_L}(q) \ . $$
Now we wish to make use of the modular transformation properties
of the theory. For $c=8n$ it transforms with $S=1$ and
$T=e^{-2 \pi i n /3}$. It is convenient to multiply $P$ with
$\eta(q)^{8n}$ to remove the phase in the $T$ transformation.
Then the function
$\hat P(q,0,\ldots,0)=(\eta(q))^{8n} P(q,0,\ldots,0)$ transforms
as a modular function of weight $4n$.
Furthermore we know the transformation properties
of the Kac-Moody characters \KaPe\
$$ \eqalign {\tau \rightarrow \tau+1 \ &: \ \ \
  {\cal X}_i(\tau+1,\vec F)=e^{2\pi i (h_i - c/24)  }
   {\cal X}_i(\tau,\vec F) \cr
 \tau \rightarrow -{1\over \tau} \ &: \ \ \
  {\cal X}_i(-{1\over\tau}, {\vec F\over\tau})=
e^{-{i\over 8 \pi  \tau}
{k\over g} \Tr_{\rm adj} F^2 }S_{ij}
{\cal X}_j(\tau,\vec F)\ , \cr }\eqn\Trafo$$
where
$$\X_i(\tau,\vec F)=\Tr_i e^{\vec F \cdot \vec J_0}
                    e^{2\pi i \tau (L_0 -c/24)}\ , \eqn\Character$$
with the trace evaluated over the positive norm states of the
representation ``$i$''. In \Trafo\
$g$ is the dual Coxeter number of the Kac-Moody algebra, and we
have traded $q$ for $\tau$, with $q=e^{2\pi i \tau}$. The trace
in \Trafo\ is evaluated in the adjoint
representation\rlap,\foot{Conventions: $J_0^a$ is Hermitian,
$f_{abc}f_{abe}=2g\delta_{ce}$. A factor ${i/2\pi}$ in the usual
definition of Chern characters has been absorbed in $F$.}except
for $U(1)$ factors, where one can use any non-trivial
representation, provided that $k/g$ is replaced by some
normalization $N$\rlap.\foot{The natural choice is the charge-1
representation, where ``charge'' is defined to be the eigenvalue
of the spin-1 operator ``$\partial X$", \ie\ the ``lattice momentum''.
In that case $N=1$.}This
normalization turns out to be irrelevant for
our purposes.

Using \Trafo\ and the fact that the $\hat P$ must be a
modular function for $\vec F=0$, we can derive how it must transform
when $\vec F\not=0$. (This is precisely the same argument as was
used in
\ScW\ to derive the transformation properties of the chiral partition
function of heterotic strings, from which one can derive the
Green-Schwarz factorization of the anomaly.) One finds
$$ \hat P\left({a\tau+b\over c\tau+d},{\vec F\over c\tau+d}\right) =
  \exp\left[{-ic\over 8 \pi (c\tau+d)} {\cal F}^2
           \right]\  (c\tau+d)^{4n}
                \hat P(\tau,\vec F)  \ ,\eqn\TrafoTwo  $$
where we have defined
$$ {\cal F}^2 = \sum_{\ell} {k_{\ell}\over g_{\ell}}
\Tr_{\rm adj} F_{\ell}^2\ , $$
with the appropriate modifications for $U(1)$'s as explained above.

To analyse the consequences of these transformation properties we need
the Eisenstein functions,  for convenience normalized as follows
$$\eqalign{  E_2(q)&=1-24\sum_{n=1}^{\infty}
             { n   q^n \over 1 -  q^n } \ , \cr
             E_4(q)&=1+240\sum_{n=1}^{\infty}
             { n^3 q^n \over 1 -  q^n } \ , \cr
             E_6(q)&=1-504\sum_{n=1}^{\infty}
             { n^5 q^n \over 1 -  q^n } \ , \cr} $$
The last two are entire modular functions of
weight 4 and 6 respectively, whereas $E_2$ has an anomalous
term in its modular transformation
$$ E_2\left({a\tau+b\over  c\tau+d}\right)=
   (c\tau+d)^2E_2(\tau) - {6i\over\pi} c (c\tau+d) \ . $$

The anomalous term in the $E_2$ transformation can be used to cancel
the exponential prefactor in \TrafoTwo. Indeed, if we define
$$\tilde P(q,\vec F) =e^{-{1/48}E_2(q){\cal F}^2} \hat P(q,\vec F) $$
we find
$$ \tilde P\left({a\tau+b\over c\tau+d},{\vec F\over c\tau+d}\right) =
(c\tau+d)^{4n}   \tilde P(\tau,\vec F)  \ .\eqn\TrafoThree  $$
Expanding $\tilde P$ in powers of $F$ one finds that the expansion
coefficients of terms of order $m$ must be modular functions of
weight $4n+m$. To proceed, we need to know that
they are regular in the upper half-plane.

The Kac-Moody characters $\X_i(\tau,\vec F)$
are explicitly known and
regular by inspection. However, the functions $\X_{i_1,\ldots,i_L}$
are characters of an unknown conformal field theory. It can be
shown \Kir\ that these characters are regular if the chiral algebra
of this unknown theory is generated by a finite number of
currents, essentially because this limits the growth of the number
of states with increasing level. The unknown conformal field theory
is in any case unitary and rational (since the modular group closes
on the finite set of
characters $\X_{i_1,\ldots i_L}$), and all known theories
of this type have a finitely generated chiral algebra. It might
be possible to prove that this is true in general, but at present
the best we can do is assume it.
The multiplication with $\eta^{8n}$ removes the
singularity at $q=0$, so that the coefficient functions are entire
modular functions of weight $4n+m$ on the upper half-plane
{\it including} $\tau=i\infty$. Basic theorems on modular functions
can then be invoked to show that all coefficient functions must
be polynomials in $E_4$ and $E_6$.

We define the functions
${\cal E}_n$
as polynomials in $E_4$ and $E_6$ with total weight $n$.
These functions have one or more free parameters.
For example ${\cal E}_8=\alpha (E_4)^2$,
${\cal E}_{10}=\alpha E_4 E_6$,
${\cal E}_{12}=\alpha (E_4)^3 + \beta (E_6)^2$,
${\cal E}_{14}=\alpha (E_4)^2 E_6$, etc.
Clearly ${\cal E}_{12k+l}$ depends on $k+1$
parameters for $l=0,4,6,8$ and $10$, and $k$ for $l=2$.
An important linear combination of the two weight 12 functions is
$$ \Delta(q)={1\over 1728} [ (E_4)^3 -(E_6)^2 ] = (\eta(q))^{24} . $$

The characters out of which $P$ was built are traces over
exponentials of the representation matrices of each
excitation level. This yields traces of arbitrary order and over
different representations. Traces over
different representations can always be expressed in terms of traces
over some fixed representation (called the
reference representation in the following).
Furthermore all traces
can be expressed in terms of a number (equal to the rank) of basic
traces $\Tr F^{s}$, where $s$
is equal to the order of
one of the fundamental Casimir operators
of the Lie algebra (these are equal to the ``exponents'' of the
Lie algebra plus 1).
The reference representation must be chosen so that for all $s$
these basic traces are non-trivial and cannot be expressed in terms
of lower-order traces (unfortunately, this often excludes the
adjoint representation).
In the following all traces will be over the
reference representation unless a different one is explicitly
indicated.

Thus we arrive at the following expression for the character-valued
partition function
$$ P(q,F_1,\ldots,F_L)=e^{{1\over 48}E_2(q){\cal F}^2} (\eta(q))^{-8n}
  \sum_{m=0}^\infty \sum_i {\cal E}_{4n+m}(i) {\cal T}^m_i
    \ .\eqn\FirstPart $$
Here ${\cal T}^m_i$ denotes a trace of total order $m$, and $i$
labels the various traces of that order.
Such a trace has the
general form
$$ {\cal T}^m_i=\prod_{\ell=1}^L \
[\Tr(F_{\ell})^{s(\ell,i)}]^{m(\ell,i)}
\ , $$
with $\sum_{\ell} s(\ell,i)m(\ell,i)=m$, and $s(\ell,i)-1$ is one of
the exponents of the $(\ell)^{\rm th}$ Lie algebra.
Each
such trace can have a different coefficient function, as indicated
by the argument $(i)$ of ${\cal E}$.

Since the ground state is a singlet representation of the theory,
it does not contribute to any of the higher traces. This  fixes
some of the parameters in the coefficient functions ${\cal E}$.
In the
absence of the exponential ``anomaly'' factor this would simply mean
that all coefficient functions for $m>0$ must start with $q^1$ rather
than $q^0$. However, because of the extra factor this is not true
if ${\cal T}^m_i$
is a product of second-order traces. In that case
the coefficient function must
cancel the traces generated by the exponential factor multiplying the
$m=0$ terms in the sum. We can take the required terms out of
the coefficient functions ${\cal E}$ by rewriting the partition
function in the following way
$$ \eqalign{
P(q,F) &= \exp\left({1\over  48} E_2(q) {\cal F}^2
           \right) \eta(q)^{-8n} \cr
&\times \Bigg\{ {\cal E}_{4n}(0) +  (E_4(q))^n
  \left[ \cosh\left({1\over48} \sqrt{E_4(q)} {\cal F}^2
\right) - 1\right]\cr
 &- ( E_4(q))^{n-3/2}  E_6(q)
  \left[ \sinh\left({1\over48} \sqrt{E_4(q)} {\cal F}^2\right)\right] \cr
 &+\sum_{m=2}^{\infty}\sum_i \Delta{\cal E}_{4n+m-12}(i)
  {\cal T}^m_i\ \Bigg\}  \ ,
 \cr  }\eqn\Traces$$
where ${\cal E}_{4n}(0)$ has  a leading term equal to 1.
Note that the cosh and sinh terms, when expanded in $F$, produce
coefficient functions that are polynomials in $E_4$ and $E_6$ of
the correct weight.
We can take out a factor $\Delta$ from the remaining
coefficient functions, because we know that they must be proportional
to $q$. This leaves ${\cal E}_{4n+m}/\Delta$, which is an entire
modular function of weight $4n+m-12$ (since $\Delta$ has no zeroes).
The functions ${\cal E }_l$ exist only for $l=0$ and $l \geq 4$,
$l$ even.
For all other values that occur in the sum they must be
interpreted as 0.

Now consider the first excited level. Expanding \Traces\ to second
order in $F$ one gets
$$ {\cal N} + \left(15-{31\over6}n
+{{\cal N}\over 48} \right) {\cal F}^2
            + \sum_{\ell} \alpha_{\ell} \Tr_{\rm adj}
            F_{\ell}^2 \ .\eqn\Sone $$
Here $\alpha_{\ell}$ is the leading coefficient of
${\cal E}_{4n-10}(\ell)$ (times a factor for the conversion from
reference to adjoint representation).
This term vanishes if $n\leq 3$.
Since by construction the first excited level (the spin-1 currents)
consists entirely of adjoint representation of the Kac-Moody algebras,
the result should be equal to the Chern-character
$\Tr e^{F.\Lambda}$, where
$\Lambda$ is the adjoint representation matrix. Upon expansion this
yields, for non-Abelian algebras
$$ \sum_{\ell} \left( \dim_{\ell} + {1\over2} \Tr_{\rm adj}
F_{\ell}^2 \right) \ . \eqn\Stwo  $$
For $U(1)$ factors there is no
$F^2$ contribution in \Stwo, and any non-trivial representation can be
used for the other traces. Comparing \Sone\ and \Stwo\ we get,
for non-Abelian algebras
$$ \eqalign { \sum_{\ell} \dim_{\ell}  = \N~~~~{\rm and}~& \cr
 \left(30-{31\over3}n+{\N\over 24}\right) {k_{\ell} \over g_{\ell}}
                + \alpha_{\ell} &= 1 \cr }\eqn\TwoFormulas$$
For $n>3$ (\ie\ $c\geq 32$) the second equation simply determines
the coefficients $\alpha_{\ell}$, and one does not learn anything
about the possible Kac-Moody algebras. However, for $n\leq3$ these
coefficients are absent, and we get
$$ {g_{\ell}\over k_{\ell}} = 30-{31\over3} n + {\N\over  24} \ ,
\eqn\Rone $$ which is
independent of $\ell$. For $U(1)$ factors the right-hand side of
the second equation in \TwoFormulas\
is zero instead of one,
and $k_{\ell}/g_{\ell}$ is replaced
by the non-vanishing normalization constant $N_{\ell}$.
Hence in this
case we find (if $n\leq 3$)
$$\N=248n-720 \ .\eqn\Rtwo $$
This makes sense only if $n=3$. Then one finds that $\N=24$, and
substituting this into \Rone\ we conclude that any non-Abelian
factor that might still be present must have vanishing dual
Coxeter number. Since this
is not possible, all 24 spin-1
currents must generate $U(1)$'s. This saturates the central charge,
and hence
the entire theory can be written in terms of free bosons with
momenta on a Niemeier lattice. The only such lattice
with 24 spin-1 currents is the Leech lattice. Therefore this is the
only meromorphic $c=24$ theory in which Abelian factors appear.

Hence we may ignore $U(1)$'s from here on, and focus on non-Abelian
factors. One can find the solutions of \Rtwo\ by determining for
each ${\cal N}$ the allowed Kac-Moody algebras, and then trying to
combine them in such a way that the total adjoint dimension is
${\cal N}$. In addition the total Kac-Moody
central charge must be less than
$8n$.
For $n=1,2$
we get only four solutions: $E_{8,1}$ for $n=1$, and $(E_{8,1})^2$,
$D_{16,1}$, and $B_{8,1}$ for $n=2$. However, for $n=1$ and $2$ the
number of spin-1 currents is not a free parameter: it must be equal
to $248n$, which eliminates the fourth solution.

It is instructive to compute the total Kac-Moody central charge:
$$\eqalign{ c_{\rm tot} &= \sum_{\ell} {k_{\ell}  \dim_{\ell} \over
                              k_{\ell} + g_{\ell} }   \cr
     &= 24\ { {\cal N} \over 248 (3-n) + {\cal N}}\ , \cr } $$
which is valid  only if $n\leq 3$.
For $n=3$ we see that the result is always equal to 24,
which implies that
the Kac-Moody system ``covers'' the entire theory, and that the
unknown part of the theory defined above is necessarily trivial.
Our results so far
can be summarized as follows

\subsection{Theorem}
{\it Let ${\cal C}$ be a modular invariant
meromorphic $c=24$ theory whose chiral algebra is finitely generated
and contains $\N$
spin-1 currents, with ${\cal N}\not=0$. Then either $\N=24$, and
${\cal C}$ is the conformal field theory of the Leech lattice, or
$\N>24$, and the spin-1 currents form a Kac-Moody algebra with
total central charge $24$. The values of $g/k$ for each simple factor
of this algebra are equal to one another,
and given by $\N/24-1$. }

This is all that can be learned from the trace identities at the
first level. The identities for higher-order traces involve
(for $n\geq3$)
always unknown coefficients analogous to $\alpha_{\ell}$ above.
These coefficients can be determined and then used to
compute traces over the second excitation level.

To write down these higher-order trace identities we first
need some definitions. The indices $J_{m_1,\ldots,m_r}(R)$ of a
representation $R$ of a simple Lie algebra are defined as
$$ \Tr_{R} F^m = \sum J_{m_1,\ldots,m_r} (R)
\prod_{i=1}^r \Tr (F^{s_i})^{m_i}\ , $$
where the traces on the right-hand side are over the reference
representation, and $\sum_i m_i s_i=m$. Here $r$ is the rank of
the Lie algebra, and the sum is over all combinations of basic traces
with the correct total order $m$. Note that with this definition
the indices depend on the reference representation.
For our purposes
it will be sufficient to consider the coefficients
$J_{m,0,\ldots,0}$, \ie\ the coefficient of $(\Tr F^2)^{m}$.
In a tensor product of $L$ Kac-Moody algebras we will
denote the coefficient of $(\Tr (F_1)^2)^{n_1}\times \ldots\times
(\Tr (F_L)^2)^{n_L}$ for a representation $R=(R_1,\ldots,R_{\ell})$
as $K_R(n_1,\ldots,n_L)$. Thus
$$ K_R(n_1,\ldots,n_L)=\prod_{\ell=1}^{L}
J_{n_{\ell},0,\ldots,0}(R_{\ell}) \ . $$

The second-level
trace identities can now be derived from \Traces. After a rather
lengthy computation we get
$$\eqalign{
\sum_R K_R(n_1,\ldots,n_L)&=\left[ \prod_{{\ell}=1,n_{\ell}\not=0}^L
{(2n_{\ell}-1)! \over 2^{ n_{\ell}-1} (n_{\ell}-1)!}
\left({k_{\ell} \over 2 N_{\ell} }\right)^{n_{\ell}}\right] \cr
&\times\left[ C_P - \sum_{{\ell}=1}^L \sum_{k=1}^{n_{\ell}}
{ 2^{k+1} n_{\ell} ! \over (n_{\ell}-k)!(P+k-1)! B_{2k}}
\left({2N_{\ell} \over k_{\ell}}\right)^k C_{k,\ell}
\right]      \ , \cr}     \eqn\HigherTrace
$$
which is valid if the total order, $P=\sum_{\ell} n_{\ell}$, is smaller
than or equal to 5.
The
sum on the left-hand side is over all representations appearing
at the second excitation level. The identity is valid for any
(non-trivial) choice of reference representation. The dependence
on this choice enters via the exponential ``anomaly'' factor in
\Traces, and manifests itself through the normalization constants
$N_{\ell}$. They are defined by the quadratic
trace of the reference representation matrices $\Lambda_{\ell}$
in the ${\ell}^{\rm th}$ group
$$ \Tr \Lambda^a_{\ell} \Lambda^b_{\ell} =  2N_{\ell} \delta^{ab}\ . $$
If one chooses the adjoint representation one must set
$N_{\ell}=g_{\ell}$ (the adjoint is a valid choice as long as only
quadratic traces appear). The coefficients $C_{k,\ell}$ are the
indices of the adjoint representation in the ${\ell}^{\rm th}$ factor,
\eg\ $C_{k,1}=K_{\rm adj}(k,0,\ldots,0)$, with respect to the
reference representation.
The coefficients $C_L$ in \HigherTrace\ are respectively equal to
196884, 32760, 5040, 720, 96, and 12 for $P=0,1,2,3,4$, and 5, where
$P$ is the total order of the trace, $P=\sum_{\ell} n_{\ell}$.
Finally, $B_{2k}$
are the Bernoulli numbers.

There is a trace identity of order $P$
whenever the function ${\cal E}_{2P}$ has
one (or fewer) parameters.
Hence one expects also an identity for $P=7$. This one is more
subtle, since one has to cancel the undetermined parameter
of ${\cal E}_{12}$ by subtracting traces of order 12. The result is
$$ \eqalign{ 12 K(n_1,\ldots,n_L) &-
\sum_{\ell=1,n_{\ell}\not=0}^L {k_{\ell} \over 2 N_{\ell}} n_{\ell}
(2n_{\ell}-1)
 K(n_1,\ldots,n_{{\ell}-1}, n_{\ell}-1,n_{{\ell}+1},\ldots,n_{\ell})
  =              \cr
 &\times
                            \left[ \prod_{{\ell}=1,n_{\ell}\not=0}^L
{(2n_{\ell}-1)! \over 2^{ n_{\ell}-1} (n_{\ell}-1)!}
\left({k_{\ell} \over 2 N_{\ell} }\right)^{n_{\ell}}\right] \cr
&\times\left[ -8                         +
\sum_{{\ell}=1}^L \sum_{k=1}^{n_{\ell}}
{ 2^{k+1} n_{\ell}! (k-6)(k+5) \over
(n_{\ell}-k)!  (k+6)!  B_{2k} }
\left({2N_{\ell} \over k_{\ell}}\right)^l C_{l,\ell}
\right]      \ , \cr}     \eqn\HigherTraceSeven     $$
Once the correct linear combination for
the left-hand side has been determined
the expression on the right-hand side
can be derived from \HigherTrace, which
is still valid for $P=6$ and $P=7$, except that the coefficients
depend on an undetermined parameter $\alpha$.
Parametrizing ${\cal E}_{12}$ in a certain way
one gets for example $C_6={9\over4}-
\alpha$, $C_7={31\over 48}-{7\over12}\alpha $. The parameter
$\alpha$ cancels if one combines the seventh- and sixth-order traces as
indicated above\rlap.\foot{There is in fact a separate free parameter
for each distinct subtrace of order 12. The precise form of the
left-hand side of \HigherTraceSeven\ is obtained by requiring the
cancellation of all these parameters.}

As already mentioned,
these identities hold independent of the choice of
reference representation.
For example, the lowest-order trace identity reads
$$ \sum_R K_R(1,0,\ldots,0)={k_1\over 2N_1}\left[ 32760
- 24 \left( {2 N_1 \over k_1 }\right ) C_{1,1} \right ] \ . $$
If the reference representation is the adjoint representation, then
$C_{1,1}=1$
and $N_1=g_1$ and the right-hand side becomes
${k_1\over 2g_1} \left[32808 -  2 \N \right ] $; if for example
the first Kac-Moody factor
is of type $A_n$ and we choose the vector representation, then
$N={1\over2}$ and $C_{1,1}=2g_1$. Now the right-hand side is larger than
before by a factor $2g_1$, but the same is true for all indices on the
left-hand side.

For higher-order traces this independence is less manifest. The
indices $J$ can be computed by means of the symmetric invariant tensors
of the Lie algebra \Pat.
There is one such tensor for each exponent,
and they are uniquely defined, up to normalization, if one requires
that their contraction with all lower-order tensors should vanish.
For example to fourth order one has
(ignoring odd traces):
$$\eqalign{\Tr \Lambda^a \Lambda^b &=I_2(R) g^{ab}  \cr
   \Tr \Lambda^{(a}\Lambda^b \Lambda^c \Lambda^{d)}
&= I_4(R) g^{abcd} + I_{2,2}(R) g^{(ab}g^{cd)}\ , \cr }\eqn\TrFour $$
where $\Lambda^a$ is a representation matrix of an irreducible
representation $R$,
and the round brackets denote symmetrization with weight 1.
The second-rank invariant tensor
$g^{ab}$ can be chosen equal to $\delta^{ab}$ by a suitable basis choice.
In this basis
tensor $g^{abcd}$ is traceless, and is fully determined
by fixing the value of
$I_4(R)$ for one representation.
A general expression is known for $I_{2,2}$:
$$ I_{2,2}(R)={3 I_2(R)^2
\over D+2}\left[{D   \over \dim(R)}
- {1\over 6} { I_2({\rm adj}) \over I_2(R)} \right ]            \ , $$
where $D$ is the dimension of the adjoint representation.
The indices $I_2, I_{2,2}$ and $I_4$ are closely related to the
indices $J_{1,0,\ldots,0}$, $J_{2,0,\ldots,0}$, and $J_{0,1,0\ldots,0}$
(or $J_{0,0,1,0,\ldots,0}$ if there is a trace of order 3, which, for
obvious notational purposes, we will from now on assume that there is
not) defined
above, but they are not quite the same. Note in particular that
$I_{2,2}$ does not depend on a choice of a reference representation.

To compute the indices $J$ one can choose some reference representation,
contract all indices in \TrFour\ with vectors $F_a$, and then solve
for $F^2$ and $g^{abcd}F_aF_bF_cF_d$ in terms of the indices of the
reference representation. Then one can express the traces over all
other representations in terms of those of the reference representation,
and read off the indices $J$. This yields
$$\eqalign{ J_{1,0,\ldots,0}(R)&={I_2(R)\over I_2({\rm ref})}  \cr
         J_{0,1,0,\ldots,0}(R)&={I_4(R)\over I_4({\rm ref})}   \cr
         J_{2,0,\ldots,0}(R)&=I_{2,2}(R)
  \left( {I_2(R)\over I_2({\rm ref})} \right )^2 -
  I_{2,2}({\rm ref}){ I_4(R)\over I_4({\rm ref})} \ . \cr   }
\eqn\FourthOrder\ $$
The dependence on the reference representation is partly through
the normalizations $I_2({\rm ref})$, $I_4({\rm ref})$
and $(I_2({\rm ref}))^2$, which cancel as explained above. The main
complication is that the last formula contains an extra term. However,
this term is proportional to $I_4(R)$ and contains no other dependence
on $R$. Since \HigherTrace\ must hold independently of the choice
of reference representation, both terms must satisfy separate
trace identities. From this we may conclude that
the terms proportional to $I_4$ (appearing on both sides of
\HigherTrace\ because $K_R$ as well as $C_{2,\ell}$ are
modified)
must
satisfy the trace identity for $J_{0,1,0,\ldots,0}$. This can
easily be checked explicitly.
Furthermore
the term
proportional to $I_{2,2}$ must satisfy the $J_{2,0,\ldots,0}$ trace
identity even when the second term in \FourthOrder\ is omitted.

This gives us one method for computing the left-hand side
of \HigherTrace\ for $P=2$.
If the third index vanishes (which is true for all Lie algebras
except $A_n, n\geq2$) there is also a formula for $I_{2,2,2}$, which
may be used instead of (but is not equal to) $J_{3,0,\ldots,0}$ for
analogous reasons:
\def\R{{D\over\dim(R)}}
\def\I{{I_2({\rm adj})\over I_2(R)}}
$$ I_{2,2,2}(R)={15 I_2(R)^3
\over(D+2 )(D+4) } \left[ \left( \R \right)^2
-{1\over2} \I\R +{1\over 12}
\left( \I \right)^2 \right]            \ . $$
This allows us to use trace identities for $P\leq3$.

A method for computing the indices $J$ directly is to use
Chern characters.
For example in $A_n$ Lie algebras the
Chern characters of the anti-symmetric tensor representations can
be expressed easily in terms of the Chern character
of the vector representation.
Suppose the Chern characters are known for some set of representations
${\cal S}$. Now tensor each element of ${\cal S}$ with one of
the antisymmetric tensor representations. If only one new
representation appears in the product, one can compute its
Chern character and enlarge the set ${\cal S}$
(Chern characters are
multiplied for tensor products and added for direct sums). It is
easy to see that for $A_n$ this procedure will yield all
representations. For other algebras we are already able to go
to sixth-order in $F$ ($P=3$),
which turns out to be sufficient.

In these computations
one has to take into account the vanishing
relations due to the non-existence of certain
fundamental traces. For $A_n$ these are the traces of order
larger than the rank plus one. To remove them
one starts by expanding the
Chern character of the vector representation up to the required
order, an then one substitutes the vanishing relations. This yields
a polynomial involving only fundamental traces, whose
coefficients are the indices $J$ of the vector representation
(the natural choice for the reference representation). To obtain
the indices of all other representations is then a matter of
straightforward multiplication and addition of polynomials.
The vanishing relations, as well as a more detailed account of
this method, may be found in \ScW.

Our strategy is now to compute the right-hand side of \HigherTrace,
and then try to match it with the traces of some set of
irreducible representations on the left-hand side. This set
of representations consists of the descendants of the ground state
and the spin-1 states, plus a choice of the
spin-2 primary fields of a given Kac-Moody
combination.
Equation
\HigherTrace\ yields diophantine equations for
the multiplicities of these primary fields, which must be positive
integers. If these equations do not have a solution, there cannot
exist a modular invariant partition function for the combination
under consideration.

The main purpose of this method is to rule out ``fake'' solutions to
the first level conditions. These conditions (summarized in the
theorem above) yield 221 solutions. Especially for small groups
such as $SU(2)$, accidental solutions should certainly be
anticipated. One may hope to eliminate them by means of
second-level trace identities.
The main advantage of using the trace identities
instead of directly checking the conditions for modular invariance is
of course that we  only have to deal with spin-2 currents, whereas in the
latter case all integer spin currents need to be included. However,
in many cases the number of spin-2 currents is still much too large.
For example, one of the 221 combinations is $(A_{1,16})^9$. This has
$(17)^9$ primary fields, of which 581820 have spin 2. The number of
equations we have at our disposal
to determine all these variables is ``only'' 8437.
Clearly this is
still untractable. Note that since only $A_1$ Kac-Moody algebras
appear, whose only fundamental trace is the quadratic one, there is
no chance to get more equations than we already have.

The large number of primary fields in
this (and many other examples) turns
out to be due to large combinatorial factors arising from permutations
of identical Kac-Moody factors. The solution to this problem is to
sum over all permutations
of the orders $n_1,\ldots,n_L$ of the
equations
within identical Kac-Moody factors.
In the present example,
this reduces the number of equations to just 34, but it also
reduces the number of variables, since fields that are permutations
of each other become indistinguishable from the point of view of
the symmetrized equations. Note that the equations are already
insensitive to the difference between (complex) conjugation and
$SO(8)$ triality.
Conjugation
forms, together with permutation of identical factors and triality,
the group of outer automorphisms of the
Dynkin diagram of the Lie algebra.
By
symmetrizing the equations
we are thus identifying all representations
related to one another by outer automorphisms.
In the example, the number of variables is
reduced to 62. This would still be too much for
real variables, but since they are positive integers the situation
improves drastically. In this case we find that no
positive integer solution exists. Of course,
if solutions {\it do} exist, we
still have to disentangle the symmetrization.

For all 221 combinations this computation is now manageable.
The maximal number of variables that occurs is 288. Typically, the
number of equations is roughly the same or much larger than the
number of variables. A computer was used to solve these equations,
but no limit was imposed on the size of the integer coefficients
(of course there is an absolute maximum, namely 196884).
We are finally left with
69 combinations for which there are solutions to all equations
considered,
including of course the 39 known cases.

\chapter{Modular Invariance}

Since ``accidental'' solutions to all trace identities are highly
improbable, we expect modular invariant partition functions to exist
in all 69 cases. Therefore it is not worthwhile to consider trace
identities for spin-3 currents. Although, given the representation
content of the second level, the spin-3 currents have to satisfy even
more equations (traces of order up to 26 can be used), the computations
become forbiddingly complicated, and would anyhow not settle the
existence of modular invariants definitively.

In principle, the conditions for modular invariance are much simpler
than the trace identities. The partition function has the form
$$ {\cal P}(\tau)=\sum m_i {\cal X}_i(\tau)\ ,\eqn\Modinv $$
where ${\cal X}_i$ is a combination of Kac-Moody characters for the
combination of fields labelled by $i$. Invariance under $T$ implies that
$i$ must have integer spin, and invariance under $S$ that the positive
integers $m_i$ must be an eigenvector of $S$ with eigenvalue 1.
The obvious strategy for solving this is to enumerate all integer
spin fields, and then solving the set of linear
equations $\sum_i S_{fi}m_i=0$,
where $f$ is a fractional spin field.
The number of fractional spin fields is much larger than that of
integer spin ones, so that there is no lack of equations.
(In all cases considered,
the solutions $m_i$ turn out to satisfy also the remaining equations,
$\sum_i S_{ji} m_i = m_j$, where $j$ has integral spin.)

The problem with this approach is again the large number of variables
that occur in certain cases due to permutations. Now symmetrization
does not help, since this does not determine $m_i$ completely and does
not even settle the existence of a solution, only its non-existence.
The solution to this problem is to make use of simple currents.

Simple currents are primary fields $J$
whose fusion rules with any primary
field yield just one field. Obviously this organizes the set of fields
into orbits, and it also assigns charges to all fields.
One would expect the presence of a simple current in the chiral algebra
to greatly reduce the amount of work needed to
determine the rest of the algebra, since this effectively reduces
the number of primary fields by ${1\over N^2}$, where $N$ is the order
of the current. One factor of $N$ is due to the fact that fractionally
charged fields are projected out, and the second one is due to
combining $N$ fields into a single one, a primary field of an algebra
which has been extended by $J$ (this counting argument is modified
if the current has fixed points). The idea is now to consider the
spin-2 content of the theory, previously determined from the
trace identities, and check whether any of those fields are simple
currents. This knowledge can then be used to simplify the search
for modular invariants.

First, some of the previous intuitive statements have to be made more
precise. Consider thus a partition function built out of characters
of some CFT, as in \Modinv, and suppose that one of those
characters corresponds to a simple current $J$. Closure of the chiral
algebra implies that $J$, acting on any other current in the
algebra, must yield another such current. This immediately rules out
fields with fractional charge with respect to $J$, since in that case
$J$ changes the conformal weight by a fractional amount, leading to
a violation of $T$-invariance. Now we prove

\subsection{Theorem} {\it Suppose that
a simple current $J$ appears in a
modular invariant of the form \Modinv\ with multiplicity
$m_J >0$. Then $m_i$ is constant on the orbits of $J$.}

\subsection{Proof} Define $m'_i=m_{Ji}$, where $Ji$ denotes the
field obtained from $i$ by the action of $J$. In the presence of
simple currents, the matrix $S$ satisfies \ScYb\Alig
$$    S_{j,Ji}=e^{2\pi i Q(j)} \ , $$
where $Q(j)$ is the charge of the field $j$. Modular invariance
implies
$$ m_i = \sum_j S_{ij} m_j\   . $$
The summation index $j$ is equal to $Jk$ for some other field $k$,
uniquely determined by $j$. Thus we get
$$\eqalign {m_i &= \sum_k S_{i,Jk} m_{Jk} \cr
                &=e^{2\pi i Q(i) }\sum_k S_{i,k} m'_k \ . \cr } $$
Now we make use of the fact that $J$ appears in the algebra. This
implies that all other fields in the algebra must
have integer charge, \ie\ either
$m_i=0$ or $Q(i) \in \Zbf$. Hence the phase in the foregoing
equation may be omitted, and we get
$$ m'_k = \sum_i S^{-1}_{k,i} m_i = m_k \ , $$
because of modular invariance of $m$. Hence $m$ is invariant under
$J$-shifts, \ie\ $m$ must be constant on the $J$-orbits.

An immediate consequence is that the multiplicity of $J$ itself must
be equal to that of the identity, \ie\ equal to one.
(The generalization of this result to arbitrary  modular invariants
$ {\cal X}_i M_{ij} {\cal X}_j^* $ is the statement
that $M_{ij}=M_{J_Li,j}=M_{i,J_Rj}$,
if $J_L$ ($J_R$) is a current in the left (right) chiral algebra. This
was proved in a somewhat different way in \BeBT, but only in the case
where the chiral algebra exists only out of simple currents, which is
not true here.)

A similar result applies to charge conjugation. Since $S^2=C$, the
charge conjugation matrix, one gets immediately $Cm=S^2m=m$, so that
the coefficients are invariant under simultaneous charge conjugation
in all Kac-Moody factors.

The knowledge that each simple current can appear only once is usually
enough to reconstruct the set of spin-2 simple currents from the
symmetrized multiplicities. In a few complicated cases it was
helpful (though probably not necessary) to determine the spin-2
multiplicities with less than maximal symmetrization. Having
determined a set of simple-current orbits that must appear in the
chiral algebra, we eliminate all fractional charge fields, and
rewrite the equations for $m_i$ as follows
$$    \sum_{i_0} S_{f,i_0} N_{i_0} m_{i_0}   = 0  \ , $$
where $f$ is an integral charge, fractional spin field, and the sum
is over all orbits of integral charge, integral spin fields. Each
such orbit is represented by one field $i_0$, and $N_{i_0}$ is the
number of fields on an orbit. It should be emphasized that $S$ is
the original Kac-Moody modular transformation matrix, and {\it not}
the matrix of the theory with a chiral algebra extended by simple
currents. The latter is in general not easy to determine because of
fixed points.

An illustrative example is $(A_{2,3})^6$. This combination has
$10^6$ primary fields and 6819 spin-2 currents, which are permutations
of only 9 distinct primary field combinations. The unique solution
to the trace identities for the spin-2 fields is
$$ 30\times[(3,0)^2 (0,0)^4]+15\times[(1,1)^4(0,0)^2]
  +30\times[(0,2)^2 (0,1)^4]+12\times[(1,2)  (0,1)^5] \ , $$
where $(n,m)$ are  $A_2$ Dynkin labels, and the square brackets
denote a representative from a set of fields identified under
charge conjugation and permutation. The result tells us to select
30 elements of the set of 60 spin-2 simple currents (\ie\ 15 permutations
of $[(3,0)^2(0,0)^4]$ times a factor 4 from charge conjugation).
Requiring locality of the currents with respect to each other,
and using the fact that each of them can appear only once, one easily
determines the solution, which is unique up to conjugation in each
$A_{2,3}$ factor. It consists of all permutations of
$[(3,0)^2(0,0)^4]$ plus their conjugates. The simple currents generate
a $(\Zbf_3)^5$ subgroup of the center, and the ``na\"\i ve'' estimate
of the number of integral charge orbits is thus $10^6/3^{10} \approx 17$.
Because of fixed points the actual number is somewhat larger (53),
including 19 integral spin orbits. Solving the equations for these
19 variables is easy, and one finds that all occur with multiplicity
one, except the fixed point field $(1,1)^6$, which occurs with
multiplicity 6.

In many cases we find that after taking into account the simple
currents, {\it all} remaining orbits
occur in the algebra with multiplicity 1.
There are several exceptions where some orbits do not appear, and a few
with multiplicities larger than 1. This happens only for orbits that
are fixed points of the simple currents,
and can be interpreted as follows.
If one were to extend the original Kac-Moody algebras with the simple
currents, the new diagonal invariant has multiplicities $N_0/N_f$ on
its diagonal, where $N_0$ is the length of the identity orbit, and $N_f$
the length of the fixed-point orbit.
The well-known interpretation is that
the extended theory contains $N_0/N_f$ fields corresponding to this term
in the modular invariant. In the example discussed
above there are thus 243 fields in the ``intermediate'' theory
with representation $(1,1)^6$. When the algebra is extended further,
six of those fields appear in the chiral algebra. From the point of view
of the intermediate theory, these are however
distinct fields (or at least
that is a logical possibility; to check this, one has to construct the
matrix $S$ of the intermediate theory by
resolving the fixed points, which
is not an easy task). All occurrences of higher multiplicities in the
69 solutions are consistent with this interpretation.

In some
cases there are no simple currents with spin 2 (for example $C_{4,10}$
has a simple current of spin 10).
Luckily, in those cases it turned
out to be possible to determine the $m_i$'s completely without
reducing the set of primary fields by means of simple currents.

%Finally one combination required some extra work, namely
%$(A_{1,4})^{12}$. The symmetrized spin-2 solution is unique
%$$ 66  \times [4^2,0^{10}]+ 12 \times [3,1^{11}]
%  + 132 \times [2^6,0^6] \ . $$
%The 66 simple currents can only be selected in one way, namely
%all permutations of $[4^2,0^{10}]$. This set of simple currents
%takes one representative of $[3,1^{11}]$ into the 11 other permutations,
%so that this contribution is also completely fixed. The difficulty is
%to choose 132 out of the ${12 \choose 6}=7\times 132$
%permutations of $[2^6,0^6]$. Since these are not simple currents, their
%multiplicities could be higher than 1.
%
%The solution can be found by symmetrizing in
%two groups of six rather than
%one of twelve. Now there are two solutions, namely
%$$ 1\times [2^6,0^6] + 0\times [2^5,0,2,0^5] + \ldots
%0 \times [2,0^5,2^5,0] + 1\times [0^6,2^6] \ldots $$
%and
%$$ 0\times [2^6,0^6] + 6 \times [2^5,0,2,0^5] + \ldots $$
%where irrelevant terms have been omitted.
%Now it is clear that each field can appear only once, and that
%there cannot be any five-fold and single overlaps with other fields.
%Using this information we can start writing down the solution.

For all 69 solutions of the trace identities we wish to prove
existence as well as uniqueness (up to equivalence)
of the meromorphic modular invariant. No further work is needed
for one third of the solutions, for which both of these
features follow from the work of Niemeier. For the 14 theories
corresponding to $\Zbf_2$ Niemeier lattices existence has been
proved in \DGM, but this does not rule out the existence of other
partition functions for the same Kac-Moody combination.

Uniqueness holds in general only up to the outer automorphisms
described above.
Furthermore, we often find additional solutions
with spin-1 currents in the chiral algebra\rlap.\foot{In particular
all combinations with $\N=48$ have at least one meromorphic
modular invariant of this type, since one can easily show that they
can all be embedded conformally
in the $D_{24}$ Niemeier CFT.}Their
presence
implies an enlargement of the Kac-Moody algebra of the theory,
with the original theory conformally embedded in the new one. Clearly
these invariants should not be counted as separate theories, since
they will be encountered again when the enlarged Kac-Moody algebra is
studied directly. Note that conformal embeddings
will never appear as solutions to the trace identities, because
they were derived under the assumption that there are no additional
spin-1 currents.

Apart from outer automorphisms and conformal embeddings,
we have found exactly one modular invariant partition
function of the form \Modinv\ for each of the 69 combinations.

\chapter{Results}

The complete results are listed in the table.
Columns 1-3 are self-explanatory. Column 4 lists the simple current
orbits that appear in the chiral algebra. Since the simple currents
form an Abelian group under addition, it is sufficient to list a
set of generators of this group. In all cases except $D_n$, the simple
currents generate a $Z_M$ group, and the elements $J^m$ of this group
are labelled by $m$. In algebras of type $A_{n,k}$, $J$ is chosen to be
the field with Dynkin labels $(k,0,\ldots,0)$; in $E_{6,k}$ the one with
Dynkin labels $(k,0,0,0,0,0)$; $B_n, C_n,  E_7$, and $E_{8,2}$
have only one
non-trivial simple current, and $G_2, F_4, E_{8,k},k\not=2$ have none.
Finally the $D_n$ simple currents are denoted as $v,s$, or $c$
if their Dynkin labels are $k$ times those of the vector, spinor or
conjugate spinor representation. The simple currents appearing in the
case of simply laced level-1 algebras (\ie\ even self-dual lattices)
were taken from \CoS, where they were called {\it glue vectors}.
We have adopted this terminology also for the other combinations
in the table.

Column 5 contains the orbits and the multiplicity with which
each orbit appears. Note that sometimes the choice of orbit
representative may hide symmetries and other relevant features.

Some notation has been introduced to deal with permutations. By
$\{ A_1;A_2;\ldots;A_K\}$ we mean all permutations of the
entries separated by semicolons; $\{\ \}_E$ means even permutations only,
and $[\ ]$ means all cyclic permutations; $(\ )^n$ means of course that
the entry between round brackets is to be repeated $n$ times.
In principle, if there is more than one Kac-Moody factor, their simple
currents and representations are separated by commas and enclosed in
round brackets. However, these symbols are omitted when no confusion
is possible.

Finally, the last column indicates where a certain conformal field
theory or modular invariant partition function has appeared before.
(Although the twisted Niemeier theories have been listed in \DGM, this
paper does not contain the partition functions.)

A few cases require a separate discussion:

\subsection{No. 0} This theory, the ``monster module'', can of course
be described in terms of a $\Zbf_2$-twisted Leech lattice.
Apparently no explicit expression is known giving its
modular invariant partition functions in terms of simpler theories,
although
one would expect this to be possible.

\subsection{No. 1} The Leech lattice \Lee\ can be
described as a modular invariant
partition function of 24 copies of ``$D_{1}$'',
putting it on a more or less
equal footing with the other Niemeier lattices. An example of a set of
simple currents yielding the Leech lattice can be found in \LSW, although
a simpler presentation might be possible. In principle one can obtain
various representations of the Leech lattice from the trace identities.
We have investigated this by solving the equations for just one $U(1)$
factor, allowing different radii for the $U(1)$. The
trace identities are ($(2N+1)!!=1\times3\times5\times...\times 2N+1$)
$$ \sum_{q=0}^{n} M_q \left( {q\over \sqrt{2n}}\right)^{2P} =
C_P (2P-1)!! \ \ (P=0,\ldots 5) $$
and
$$ \sum_{q=0}^{n} M_q \left[ 12  \left({q\over \sqrt{2n}})\right)^{14}
-91 \left({q\over\sqrt{2n}}\right)^{12}\right] = -8 . 13!! \ , $$
where $C_P$ are the coefficients appearing in the trace identities
\HigherTrace, and $n$ defines the radius, in such a way that one
obtains a rational $U(1)$ with $2n$ primary fields with charges
$ q/\sqrt{2n}, q=-n+1,\ldots,n$. Note that opposite charges give the
same contribution to the traces, so that the sums can be reduced
to half this range.  (One can reduce them even further by requiring
that the conformal weights do not exceed 2,
\ie\ $q \leq 2\sqrt{2n}$.)
For $n=1$ these identities are not
valid because they do not take into account the charge $\pm \sqrt2$
states (the $SU(2)$ roots) appearing at the first level; likewise, for
$n=2$ there are charge $\pm 2$ ($q=4$)
descendants at the second level, which
must be included in the sum on the
left-hand side, with multiplicity
$M_4$ equal to 2 (\ie\ 1 for each charge). For larger $n$ the first
such descendant appears at level $n$, and does not affect the argument.
One may try to solve the equations to obtain the unknown multiplicities
$M_q$ of the primary fields. For $n\leq10$ we always found one or
more solutions (since the number of contributing fields increases
with $n$ it does not make much sense to consider larger values). For
example, for $n=2$ the equations are satisfied with
$M_0=93474, M_1=94208$, and $M_2 =9200$. This can be inverted to
obtain a somewhat strange closed formula for the
coefficients $C_P, P\geq 1$:
$$ C_P (2P-1)!!=92(100+2^{10-2P})+2^{2P+1} \ . $$
The fact that solutions exist for larger values of $n$ indicates that
there are many other ways to write the Leech lattice as a product of
rational $U(1)$'s. Constructing some of these by considering traces over
more than one $U(1)$ factors might be feasible; but proving that all
solutions are in fact the same, up to rotations, is certainly impossible
with these methods alone.
Fortunately, this is already known by other arguments.

\subsection{No. 2}
This was one of the most difficult cases, owing to the presence of
a large number of fixed-point fields. The solution is, with
the set of simple currents listed in the table:
$$ (0)^{12} + (1)^{11}3 + \{(2;)^9 (0;)^3\}+ 12 \times (2)^{12}
+ 132  \hbox{~chosen from } \{2^6 0^5\}  \ , $$
\ie\ all 220 permutations of the spin-3 fields $(2)^9 (0)^3$ appear once,
but only ${1\over 7}$ of the permutations of $(2)^6 (0)^6$. Using
identities symmetrized on subsets one can show that all 132 must
be different, but the hard problem is to select them.
The answer can be characterized as follows. Take the set of 11 vectors
$$     1[0;0;1;0;0;0;1;1;1;0;1]\ ,$$
and generate all $11^3$ vectors obtained by adding them modulo 3.
One then obtains 729 different vectors, each appearing with a
multiplicity 243. Now replace the non-vanishing entries of all
vectors (which are either 1 or 2) by the $SU(2)$ fixed-point field
$(2)$, and divide by 2 the multiplicities of all these fields, except
the identity. In this way one obtains 12 copies of $(2)^{12}$, all
220 permutations of $(2)^9 (0)^3$, and 132 different permutations
of $(2)^6 (0)^6$ ($2 \times (132+220+12)+1=729$). This is the
solution, up to permutations of the $SU(2)$ factors. The description
given here was obtained by working out the $\Zbf_2$ twist of the
$(A_2)^{12}$ Niemeier lattice\rlap.\foot{The spin-2 field $(1)^{11}3$,
12-fold degenerate because of the simple current action on it, plays
the r\^ole of one of the twist fields, referred to as $\sigma_E$ in the
introduction.}
However, this was only used to obtain
a presentable description of the 132 spin-2 fields. Uniqueness was
proved by solving the modular invariance conditions, as in all other
cases.

\subsection{No. 5}
As in the previous case, the only complication is the determination
of combinations of $SU(2)$ fixed points. The answer is
$$ (0)^{16}+8\times (1)^{16} + J {\cal S}\ , $$
where ${\cal S}$ is the set of 30 vectors
$$ n_0 (1)^{16} +
 n_1 (0)^8(1)^8 + n_2 ((0)^4(1)^4)^2 +
 n_3 ((0)^2(1)^2))^4 + n_4 (01)^8   \mod 2\ , $$
with $n_i$ defined modulo 2,
with at least one of the $n_i$, $1\geq i \geq 4$
equal to 1. As indicated, the vector entries (which of course are
$SU(2)$ Dynkin labels) are added modulo 2. The combination
$(0)^8(1)^8$ has spin ${3\over2}$. A spin-2 field is obtained by acting
once with the $SU(2)$ simple current $J$ on one of the identity
components. The simple currents in the chiral algebra imply that it does
not matter in which factor one performs this action.

\subsection{No. 12} The notation in
column 5 uses a double cyclic permutation,
which may be confusing. It means to perform all cyclic permutations of
the six representations, combined with all
permutations of the last two {\it pairs} of representations,
with all the distinct combinations that are obtained counted once.
The total number of representations obtained this way is 30.

The other modular invariants do not require further explanation.

An interesting by-product of these results is a list of some new
invariants of simple Kac-Moody algebras. They can be read off from the
table by looking for cases where non-trivial primary fields appear in
combination with the identity of all other simple Kac-Moody algebras
(if any) in
the theory. In this way we find modular invariants (with extension
of the chiral algebra with spins 2 and higher) for
$D_{4,12}$,
$D_{5,8}$,
$A_{7,4}$, $D_{5,4}$, $D_{6,5}$, $C_{5,3}$, $A_{8,3}$,
$D_{7,3}$, $C_{7,2}$, $A_{9,2}$, $C_{10,1}$,
$C_{4,10}$, $A_{5,6}$, $A_{6,7}$,
$D_{9,2}$, $B_{12,2}$,
$F_{4,6}$, $E_{6,4}$, and $E_{7,3}$.
These invariants correspond neither
to conformal embeddings, nor are they simple current invariants.
However, the first 11 are presumably\foot{This has not been
checked, but is conjectured here on the basis of the existence
of a rank-level dual conformal embedding, using the duality
relations $C_{n,k} \leftrightarrow C_{k,n}$, $SO(n)_k \leftrightarrow
SO(k)_n$, and $SU(n)_k \leftrightarrow SU(k)_n$.}
related to conformal embeddings by
rank-level duality.
The invariants to which they are related are
respectively those of $D_{6,8}$, $D_{4,10}$, $A_{3,8}\equiv D_{3,8}$,
$D_{2,10}\equiv (A_{1,10})^2$, $B_{2,12}$, $C_{3,5}$, $A_{2,9}$,
$B_{1,14}\equiv A_{1,28}$, $C_{2,7}\equiv B_{2,7}$, $A_{1,10}$ and
$ C_{1,10}\equiv A_{1,10}$. Of the remaining ones, $C_{4,10}$
is dual to $C_{10,4}$, which cannot be conformally embedded in any
Kac-Moody algebra. Hence we anticipate the existence of a higher
spin extension of $C_{10,4}$. The algebras $A_{5,6}$ and $A_{6,7}$
are ``self-dual'', and
$D_{9,2}$ and $B_{12,2}$ are formally related to
$SO(2)$, but since this is Abelian it is difficult to give meaning
to the level. Finally, the three exceptional algebras are not
dual to anything. Some of these Kac-Moody invariants have already
been obtained or conjectured on the basis of rank-level duality,
and the $F_{4,6}$ invariant appeared first in \ScYA.

\chapter{Discussion}
In this paper we have shown that all meromorphic $c=24$ conformal
field theories with finitely generated chiral algebras containing
at least one spin-1 current have a partition function which can
be written entirely as a modular invariant combination of
Kac-Moody character. Furthermore we have enumerated all 69
such partition functions, 30 of which were not yet known.
The actual construction of the
conformal field theories corresponding to these new partition
functions remains to be done. If there is exactly one CFT per
modular invariant, and only one theory without spin-1 currents, then
the total number of $c=24$ meromorphic CFT's is 71 (an interesting
fact, though perhaps a meaningless coincidence is that 71 is
precisely the largest prime in the order of the monster group).

We hope that these theories have an interesting
r\^ole to play in physics or mathematics, but this
remains to be elucidated.
The list itself could have revealed some
underlying structure, but if it exists it must be rather subtle.
In the spirit of generalizing from lattices to conformal field
theories there are several questions that suggest themselves. For
example, for self-dual lattices
a formula exists \Sie\ for the sum of the inverse orders
of the lattice automorphism groups (this is known as the
Minkowsky-Siegel formula). If there is a generalization of this
formula to conformal field theory one could use it to prove
completeness of our list, and at the same time prove uniqueness
of the monster module. A second interesting fact about the
Niemeier lattices is that they can all be embedded into the
unique Lorentzian self-dual lattice $\Gamma_{25,1}$, and are
``orthogonal'' to certain lightlike vectors on this lattice. It would
be very interesting to see if this fact has a generalization to
conformal field theory.

The last two points are pure speculation, but
in any case the list has enabled us to make some modest progress
in two other classification problems, namely string theory
in 10 dimensions, and modular invariants of Kac-Moody algebras.
The main lesson learned about the latter classification problem is
that we still know essentially nothing about it. Several new
invariants were found that could not have been anticipated with any
known method.
One might hope that this exhausts the list of exceptional
extensions of the
chiral algebra of {\it simple}
Kac-Moody algebras, but there is not really any good reason to believe
that. The situation is much worse for semi-simple Kac-Moody algebras. A
very large number of exceptional
invariants for such algebras can be read off from
the table. It would be very strange indeed if no new exceptional
invariants appear on a list of $c=32$ (or larger) conformal field
theories, which undoubtedly will never by enumerated.

Indeed, the present classification has made it clear once more that
something changes drastically beyond $c=24$ (this can also be seen in
other ways, \eg\ from the Minkowsky-Siegel formula, or from the
properties of $\Gamma_{8n+1,1}$ Lorentzian self-dual lattices).
Even though
the number of meromorphic conformal field theories is to large to allow
a complete listing (already the number of lattices is much larger
the $8\times 10^7$), one would at least like to have a finite
algorithm that can produce the list in principle.
Even for the subclass that has a $c=32$ spin-1 algebra the methods
we used for $c=24$ do not yield a finite algorithm, since one
has to allow {\it a priori} $U(1)$ factors with arbitrary radii.
These difficulties are closely related to the unsolved
problem of arriving at a practical classification of rational
conformal field theories.

\ack I would like to thank Elias Kiritsis and Hermann Nicolai
for discussions
of several aspects of this work. Furthermore I would like
J\"urgen Fuchs for making some of his computer programs
available to me, and for providing some $S$-matrices of large-rank
algebras, computed with the method explained in \DrFu.

\endpage
\input tables
\line{Table Caption\hfill}
\vskip 1.truecm
Modular invariant partition functions of the 69 theories with a
non-Abelian spin-1 algebra. The monster module and the Leech lattice
are included for completeness as Nos. 0 and 1. The table is explained
in section 4.    \vfill \break

\tenpoint
\vskip .5truecm
\begintable
No. |$\N$ | Spin-1 algebra | Glue | Orbits | Ref. \crthick
0 | 0   |     ---        |      |        |  \LFM\ \cr
1 | 24  |   $U(1)^{24}$   |      |   (0)  | \Lee\ \cr
2 | 36  | $(A_{1,4})^{12}$  | $ 1 [1;(0;)^{10}] $ | see text | \DGM\ \cr
3 | 36  | $D_{4,12}A_{2,6}$| $(0,1)+(s,0)+(v,0)$ |
 $(0000+0006+0060+0066+0400+3033,00)$ |  \nr | | | |
 $+(0204+0240+0300+0244+1411+2122,03)$ |  \nr | | | |
 $+(0044+0600+1213+1231+1233+2022,11)$ | \nr  | | | |
 $+(0004+0040+0048+0320+0302+0324 $    | \nr  | | | |
 $+1033+1035+1053+3\times2222,22)$ |      \cr
4 | 36  | $C_{4,10}$       | $1$ |
  $ 0000+0024+0040+0044+00,\!10,\!0+0260+0321$  | \nr | | | |
  $+0323+0500+0800+1051+1430+1431+2\times2222 $  | \nr | | | |
  $+2242+3031+4140$  | \cr
5 | 48  | $(A_{1,2})^{16}$ | $ 11[11;(00;)^6]$  | see text | \DGM\ \nr
 |  |                | $+1010[1010;(0000;)^2]$ |        |  \nr
 |  |                |   $+(1000)^4 $ |        |  \cr
6 | 48 | $(A_{2,3})^6$ | $1[1;(0;)^4]$ | $(00)^6 +\{(11;)^4(00;)^2\} +
               (01)^5(12)+(10)^5(21)+6\times(1,1)^6$ |  \cr
7 | 48 | $(A_{3,4})^3 A_{1,2}$ | $ [1;0;0]1$ |
$((000)^3+(012)^3,0)+ (\{002;010;111\},1) + 4\times ((111)^3,1)$ |
\nr | | | |
$+([000;020;020],2)+([012;020;020],2) $ |  \cr
8 | 48 | $A_{5,6}C_{2,3}A_{1,2}$ | $(1,0,1)+(0,1,1)$ |
$ (00000+02020,00,0)+
  (00003+00211,30,1) $ | \nr | | | |
$+(00200+02020,20,2)+
  (00130+03100,11,1) $ | \nr | | | |
$+(00022,01,0)+
  (00030,00,2)+
  (01102,10,1)  $ | \nr | | | |
$+(01121,20,0)+
  (01210,01,2)+
  2\times (11111,11,1) $ |  \cr
9  | 48 | $(A_{4,5})^2$ | $(1,0)+(0,1)$  |
 $(0000,0000)+
  (0102,0102)+
  4\times(1111,1111) $ | \nr | | | |
 $+[0021;0110]+
  [1111;0013]$ | \cr
10 | 48  | $D_{5,8}A_{1,2}$ | $(s,0)$ |
 $(00000+00222+03011+10111,0)+$ | \nr | | | |
 $(00113+00131+2\times11111,1)+$ | \nr | | | |
 $(00044+00200+01022+01211,2)$ | \cr
11 | 48  | $A_{6,7}$      | $1$  |
  $000000+001301+103100+002030$ | \nr | | | |
  $ +010122+ 3\times111111$  |\cr
12 | 60 | $(C_{2,2})^{6}$ | $1[1;(0;)^5]$ |
$(00)^6+[00;(20;)^5]+[00;(01;)^5]+[11;(10;)^5] $ | \DGM\ \nr | | | |
$+\{(01;)^3(20;)^3\}+[(00;)^2[(01;)^2;(20;)^2]]$ | \cr
13 | 60 | $D_{4,4}(A_{2,2})^4 $ | $(0,1,1,1,0)$ |
$ (0000,(00)^4)+
  (0100,(11)^4) $ | \nr | | | $+(0,2,1,0,1) $ |
$+(0200,[11;(00;)^3])+
  (1011,[00;(11;)^3])$ |  \nr
   |    |                       | $+(v,0,0,0,0) $  |
$+(0002,00,11,11,00)+
  (0002,11,00,00,11) $ | \nr | | |$+(s,0,0,0,0)$  |
$+(0020,00,00,11,11) +
  (0020,11,11,00,00) $ | \nr | | |      |
$+(0022,00,11,00,11) +
  (0022,11,00,11,00) $ | %cr
\endtable
\begintable
No. |$\N$ | Spin-1 algebra | Glue | Orbits | Ref. \crthick
14  | 60 | $ F_{4,6} A_{2,2}$ | --- |
$ (0000+0004+0030+1100,00) $ | \ScYc\  \nr
    |    |                  |     |
$ +(0003+0006+0021+2010,11) $ | \nr
    |    |                  |     |
$ +(0101+1012,10+01)+(0102+2000,02+20)$ |
 \cr
15 | 72 | $(A_{1,1})^{24}$ |
$1[(0;)^5 1;0;1;(0;0;1;1;)^2$ | (0) | \Nie\  \nr | | |
$~~~~~~~0;1;0;(1;)^4]$ |     |
\cr
16 | 72 | $(A_{3;2})^4(A_{1,1})^4$ |
$(1100)^2 + (1010)^2 $  |
$((000)^4 + [000;(101;)^3],(0)^4) $  |\DGM\ \nr | | |
$+ (1001)^2  +(2,(0)^3,(1)^4)$ |
$+ ((001)^3,011,0,1,0,0)$ |       \cr
17 | 72 | $A_{5,3}D_{4,3}(A_{1,1})^3 $ |
$(0,s,0,1,1)+(0,v,1,1,0) $ |
$  ((00000,0000)+(00111,1011)+(01002,0100),(0)^3)  $ | \nr | | |
$+(1,0,1,1,1)$ |
$  +((01010,0002+0020+2000,(0)^3) $ | \cr
18 | 72 | $A_{7,4}(A_{1,1})^3$ | $(1,1,0,0)$ |
$ (0000000+0001101,(0)^3)$ | \nr | | | |
$+(0000202+0010100,0,1,1)$| \nr | | | |
$+(0010011,0,0,1)+(0010011,0,1,0)$ | \cr
19 | 72 | $D_{5,4}C_{3,2}(A_{1,1})^2 $ |
$(0,1,1,1)+(s,1,0,0)$ |
$ ((00000,000)+(00200,000)+(01011,020),0,0)$ | \nr | | | |
$+((00011,010)+(00022,020)+(00200,010),1,1)$ | \nr | | | |
$+((00100,110)+(00111,001)+(10011,011),0,1)$ | \cr
20 | 72 | $D_{6,5}(A_{1,1})^2$ | $(s,0,1)+(c,1,0)$ |
$ (000000+010002+010020,0,0)$ | \nr | | | |
$+(100111+002000+200100,0,0) $ | \cr
21 | 72 | $C_{5,3}G_{2,2}A_{1,1}$ | $(1,0,1)$  |
$ (00000+00020+03000,00,0) $ | \nr | | | |
$+(00002+02000+10110,02,0) $ | \nr | | | |
$+(10001+00030+02010,10,0) $ | \nr | | | |
$+(00200+01101+20010,01,0) $ | \cr
22 | 84 | $C_{4,2}(A_{4,2})^2$ | $(1,0,0)+(0,1,2)$ |
$ (0000,(0000)^2)
 +(0020,(0110)^2)
 +(0100,(1001)^2) $ | \nr | | | |
$+(0001,[0000;0110])
 +(0200,[0000;1001]) $ | \nr | | | |
$+(1010,[0110;1001]) $  |  \cr
23 | 84 | $(B_{3,2})^4$ | $ 1[1;0;0] $   |
$(000)^4+(001)^3(101)+\{000;010;100;002\}_{\rm E} $ | \DGM\ \nr | | | |
$+[100;(002;)^3]+[002;(010;)^3]+[010;(100;)^3] $ |  \cr
24 | 96 | $(A_{2,1})^{12}$ |$2[1;1;2;(1;)^3(2;)^3 1;2]$| (0) |
\Nie\ \cr
25 | 96 | $(D_{4,2})^2(C_{2,1})^4$ |
$([s;0],1,1,0,0)$ |
$ ((0000)^2,(00)^4)
 + ([0000;0100],(10)^4)
                        $ | \DGM\ \nr | | |
$+([v,0],0,1,1,0)$ |
$+  ((0001)^2,00,01,10,10)
 + ((0001)^2,10,10,00,01)$ | \nr | | |
$+ (0,0,1,1,1,1) $ |
$ +  ((0010)^2,00,10,01,10)
+  ((0010)^2,10,00,10,01) $ | \nr | | | |
$ +  ((0011)^2,00,10,10,01)
+  ((0011)^2,10,00,01,10) $ | \nr | | | |
$+ ((0100)^2,(00)^3,01) $ | \cr
26 | 96 | $(A_{5,2})^2C_{2,1}(A_{2,1})^2) $ |
$(1,0,1,1,1)+(0,1,1,1,2)$ |
$ ((00000)^2,(00)^3)
 +((01010)^2,01,(00)^2) $ | \nr | | | |
$+([00001;10010],10,01,00)  $  | \endtable
\begintable
No. |$\N$ | Spin-1 algebra | Glue | Orbits | Ref. \crthick
27 | 96 | $A_{8,3}( A_{2,1})^2$ | $ (1,1,1)$   |
$ (00000000,(00)^2)
 +(00010101,[01;10])
 +(00100011,(00)^2) $  | \cr
28 | 96 | $E_{6,4} C_{2,1} A_{2,1} $ | $(1,0,1)$ |
$ (000000+100011,00,00)
 +(110000+000110,10,00) $ | \nr | | | |
$ (200020+001000,01,00)
                 $ | \cr
29 | 108 | $(B_{4,2})^3 $ | $1[1;0]$ |
$ (0000)^3
 +(0001,0001,1001)
 +(0010,0010,0010) $ | \DGM\ \nr | | | |
$+[0000;(0010;)^2]
 +[0100;(0002;)^2] $ | \nr | | | |
$+[0002;(1000;)^2]
 +[1000;(0100;)^2] $ | \cr
30 | 120 | $(A_{3,1})^8 $ |
$3[2;0;0;1;0;1;1]$ | (0) | \Nie\ \cr
31 | 120 | $(D_{5,2})^2 (A_{3,1})^2$ |
$(0,s,1,1)$ |
$ ((00000)^2+[00011;01000],(000)^2)$ | \DGM\ \nr | | |
$+(s,0,3,1)$ |
$ +((00001)^2,010,001)$ |    \cr
32 | 120 | $E_{6,3} (G_{2,1})^3 $ | $(1,0,0,0)$ |
$  (000000,(00)^3)
 + (000001,(01)^3) $ | \nr | | | |
$+  (001000,[00;00;01])
 +  (100010,[00;01;01])  $ | \cr
33 | 120 | $A_{7,2} (C_{3,1})^2 A_{3,1} $ |
$(1,0,1,1)$ |
$ (0000000,(000)^2,000)
 +(0010100,(010)^2,000) $ | \nr | | |
$+(0,1,1,2)$ |
$+(0100010,[000;010],000) $ | \cr
34 | 120 | $D_{7,3} A_{3,1} G_{2,1}$ | $(s,1,0)$ |
$ (0000000+1000100,000,00)
 +(0000011+1010000,000,01)$  | \cr
35 | 120 | $C_{7,2} A_{3,1} $ |  $(1,2)$ |
$ (0000000+0000200+1000001+0101000,000)
      $ | \nr | | | |
$+(0010010,001+100) $ | \cr
36 | 132 | $A_{8,2} F_{4,2}$ | $(3,0)$ |
$ (00000000,0000)
 +(00011000,1000)
 +(10000001,0002) $ | \nr | | | |
$+(00100100,0010)
 +(01000010,0001)$ | \cr
37 | 144 | $(A_{4,1})^6$ | $1[0;1;4;4;1]$ |  (0)  | \Nie\ \cr
38 | 144 | $(C_{4,1})^4$ | $1[1;0;0]$ |
$ (0000)^4
 +[0000;(0100;)^3]
 +((0010)^3,1000)    $ | \DGM\         \cr
39 | 144 | $D_{6,2}C_{4,1}(B_{3,1})^2 $ |
$(s,0,0,1)$ |
$ (000000,0000,000,000)
 +(000100,0100,000,000) $ | \DGM\ \nr | | |
$+(0,1,1,1)$ |
$+(000011,0100,001,001)
 +(001000,0000,001,001) $ | \nr | | |
$+(v,0,1,1)$ |
$+(000001,0010,000,001)
 +(000010,0010,001,000) $ | \cr
40 | 144 | $A_{9,2} A_{4,1} B_{3,1} $ | $(1,2,1)$ |
$(000000000+001000100,0000,000)+(001001000,0001,001)$ | \cr
41 | 156 | $(B_{6,2})^2$ | $(1,1)$ |
$ (000000,000000)
 +(000001,100001)$ | \DGM\ \nr  | | | |
$+[000010;100000]
 +[000100;000002]
 +[001000;010000]$ | \cr
42 | 168 | $(D_{4,1})^6$ | $(s)^6 +0[0;v;c;c;v]$ | (0) | \Nie\ \cr
43 | 168 | $(A_{5,1})^4 D_{4,1}$ | $2[0;2;4]0$ | (0) | \Nie\ \nr
 | | |  $+(3,3,0,0,s)$ |  |
\nr | | | $+(3,0,3,0,v)$ | | \nr | | |
 $+(3,0,0,3,c) $ | | \cr
44 | 168 | $E_{6,2}C_{5,1}A_{5,1}$ | $(1,1,1)$ |
$ ((000000,00000)+(000001,00010)+(100010,01000),00000) $ | \cr
45 | 168 | $E_{7,3} A_{5,1} $ | $(1,3)$ |
$(0000000+0000011,00000)+
 (0000100,00010+01000) $ | \endtable
\begintable
No. |$\N$ | Spin-1 algebra | Glue | Orbits | Ref. \crthick
46 | 192 | $(A_{6,1})^4$ | $1[2;1;6]$ | (0) | \Nie\ \cr
47 | 192 | $D_{8,2} (B_{4,1})^2$ | $(s,0,0)+(v,1,1)$ |
$ (00000000,(0000)^2)
 + (00000100,(0001)^2) $ | \DGM\ \nr | | | |
$ +(00000001,[1000;0001])
  +(00010000,0000,1000) $ | \cr
48 | 192 | $(C_{6,1})^2 B_{4,1}$ | $(1,0,1)+(0,1,1)$ |
$ ((000000)^2,0000)
 + ((000100)^2,1000) $  | \nr | | | |
$+ ([000010;001000],0001)$ | \cr
49 | 216 | $(A_{7,1})^2 (D_{5,1})^2 $ | $(1,1,s,v)+(1,7,v,s)$ | (0) |
\Nie\ \cr
50 | 216 | $D_{9,2}A_{7,1}$ | $(s,2)$ |
$ (000000000+000001000,0000000)
 +(000000001,0000100)$ | \DGM\ \cr
51 | 240 | $(A_{8,1})^3$ | $[1;1;4]$ | (0) | \Nie\ \cr
52 | 240 | $C_{8,1}(F_{4,1})^2 $ | $(1,0,0)$ |
$ (00000000,(0000)^2)
 +(00000100,(0001)^2) $ | \nr | | | |
$+(00010000,[0000;0001])$ | \cr
53 | 240 | $E_{7,2}B_{5,1}F_{4,1}$ | $(1,1,0)$ |
$ (0000000,00000,0000)
 +(0000001,00001,0000) $ | \nr | | | |
$+(0000010,00001,0001)
 +(0000100,00000,0001) $ |  \cr
54 | 264 | $(D_{6,1})^4$ | $\{0;s;v;c\}_{\rm E}$ | (0) | \Nie\ \cr
55 | 264 | $(A_{9,1})^2D_{6,1}$ | $(2,4,0)$ | (0) | \Nie\ \nr | | |
$+(5,0,s)+(0,5,c)$ |  |
  \cr
56 | 288 | $C_{10,1}B_{6,1}$ | $(1,1)$ |
$ (0000000000+0000010000,000000)
+ (0000001000,000001) $ |              \cr
57 | 300 | $B_{12,2}$ | --- |
$000000000000
+100000000001  $ | \DGM\    \nr | | | |
$+000000000100
 +000010000000  $ |              \cr
58 | 312 | $(E_{6,1})^4 $ | $1[0;1;2]$ | (0) | \Nie\ \cr
59 | 312 | $A_{11,1}D_{7,1}E_{6,1}$ | $(1,s,1)$ | (0) | \Nie\ \cr
60 | 336 | $(A_{12,1})^2$ | $(1,5)$ | (0) | \Nie\ \cr
61 | 360 | $(D_{8,1})^3$  | $[s;v;v]$ | (0) | \Nie\ \cr
62 | 384 | $E_{8,2} B_{8,1}$ | $(1,1)$ |
$(00000000,00000000)+(10000000,00000001)$ | \ScYc\ \cr
63 | 408 | $A_{15,1}D_{9,1}$ | $(2,s)$ | (0) | \Nie\ \cr
64 | 456 | $D_{10,1}(E_{7,1})^2$ | $(s,1,0)+(c,0,1)$ | (0) | \Nie\ \cr
65 | 456 | $A_{17,1}E_{7,1}$ | $(3,1)$ | (0) | \Nie\ \cr
66 | 552 | $(D_{12,1})^2$ | $[s;v]$ | (0) | \Nie\ \cr
67 | 624 | $A_{24,1}$ | $(5)$ | (0) | \Nie\ \cr
68 | 744 | $(E_{8,1})^3$ | --- | (0) | \Nie\ \cr
69 | 744 | $D_{16,1}E_{8,1}$ | $(s,0)$ | (0) | \Nie\ \cr
70 | 1128 | $D_{24,1}$ | $(s)$ | (0) | \Nie\ \endtable  \vfill\break
\twelvepoint
\par \penalty-4000\vskip\chapterskip
   \spacecheck\referenceminspace \immediate\closeout\referencewrite
   \referenceopenfalse
   \line{\fourteenrm\hfil REFERENCES\hfil}\vskip\headskip
   \endlinechar=-1
   \input referenc.texauxil
   \endlinechar=13
   
\end